\documentclass[12pt]{article}
\usepackage{amsthm}
\usepackage{amsfonts}
\usepackage{amssymb}
\usepackage{amsmath}

\newcommand{\D}{\mathfrak {D}}

\def\d{{\rm d}}
\def\pn{\par\noindent}

\def\v{\varepsilon}
\def\la{\langle}
\def\ra{\rangle}

\newtheorem{prop}{Proposition}[section]
\newtheorem{corr}[prop]{Corollary}
\newtheorem{theo}[prop]{Theorem}
\newtheorem{lem}[prop]{Lemma}
\newtheorem{defi}[prop]{Definition}
\newtheorem{rema}[prop]{Remark}

\newcommand{\er}{{\mathbb R}} % Reels
\newcommand{\C}{{\mathbb C}} % Complexes
\newcommand{\zed}{{\mathbb Z}} % Entiers relatifs
\def\1{1\hskip-0.09cm{\rm l}}

\begin{document}

\title{Quantum electrodynamics\\ of relativistic bound states with cutoffs.}
\author{
{\bf Jean-Marie Barbaroux\footnote{E-mail:
barbarou@cpt.univ-mrs.fr}}\\ {\small \it  Centre de Physique
Th\'eorique, Campus de Luminy}\\ {\small \it Case 907, 13288
Marseille Cedex~9, France}\\ \and {\bf Mouez
Dimassi\footnote{E-mail: dimassi@math.univ-paris13.fr}}\\ {\small
\it CNRS-UMR 7539, D\'epartement de Math\'ematiques}\\ {\small
Institut Galil\'ee, Universit\'e Paris-Nord} \\ {\small \it 93430
Villetanneuse, France} \and {\bf Jean-Claude
Guillot\footnote{E-mail: guillot@cmapx.polytechnique.fr}}\\
{\small \it CNRS-UMR 7539, D\'epartement de Math\'ematiques}\\
{\small Institut Galil\'ee, Universit\'e Paris-Nord} \\ {\small
\it 93430 Villetanneuse, France}\\ {\small \it and}\\ {\small \it
CNRS-UMR 7641, Centre de Math\'ematiques Appliqu\'ees, Ecole
Polytechnique}\\ {\small \it 91128 Palaiseau Cedex, France} }
\date{{\small November 2003}}
\maketitle

\begin{abstract}
We consider an Hamiltonian with ultraviolet and infrared cutoffs,
describing the interaction of relativistic electrons and positrons
in the Coulomb potential with photons in Coulomb gauge. The
interaction includes both interaction of the current density with
transversal photons and the Coulomb interaction of charge density
with itself.

We prove that the Hamiltonian is self-adjoint and has a ground state
for sufficiently small coupling constants.
\end{abstract}

\maketitle

\newpage

%%%%%%%%%%%%%%%%%%%%%%%%%%%%%%%%%%%%%%%%%%%%%%%%%%%%%%%%%%%%%%%%%%%%%%%%%
%%%%%%%%%%%%%%%%  INTRODUCTION %%%%%%%%%%%%%%%%%%%%%%%%%%%%%%%%%%%%%%%%%%
%%%%%%%%%%%%%%%%%%%%%%%%%%%%%%%%%%%%%%%%%%%%%%%%%%%%%%%%%%%%%%%%%%%%%%%%%

\section{Introduction}\label{S1}
In Refs. \cite{ref1} an Hamiltonian with cutoffs describing
relativistic electrons and po\-si\-trons in a Coulomb potential
interacting with transversal photons in Coulomb gauge is
considered. In that note \cite{ref1}, results concerning the
self-adjointness of the Hamiltonian and the existence of a ground
state for it are announced.

In this article, we consider the full model of QED by adding the
Coulomb interaction of the charge density with itself to the
Hamiltonian described in \cite{ref1}. We thus study an Hamiltonian
with ultraviolet and infrared cutoff functions with respect to the
momenta of photons, but also with respect to the momenta of electrons
and positrons. The total Hamiltonian in the Fock space of electrons,
positrons and photons is then well defined in the Furry picture.

In this paper, we prove results concerning the self-adjointness of the
Hamiltonian and the existence of a ground state when the coupling
constants are sufficiently small.

In \cite{ref3}, Bach, Fr\"ohlich and Sigal proved the existence of
a ground state for the Pauli-Fierz Hamiltonian with an ultraviolet
cutoff for photons, and for sufficiently small values of the fine
structure constant, without introducing an infrared cutoff. Their
result has been extended by Griesemer, Lieb and Loss \cite{ref11}
under the binding condition, and finally by Barbaroux, Chen and
Vugalter \cite{BCV}, and Lieb and Loss \cite{LL}. For related
results see \cite{ref12, ref13, ref14, ref15, ref16}.

No-pair Hamiltonians for relativistic electrons in QED have been
recently considered in \cite{BM, ref17, ref18, ref19}. In
\cite{arai2}, Arai has analyzed the hamiltonian of a Dirac particle
interacting with the quantum radiation field. In \cite{SSZ}, an
Hamiltonian without conservation of the particle number is studied. In
\cite{ammari}, the scattering theory for the spin fermion model is
studied.

The case of relativistic electrons in classical magnetic fields was
studied earlier in \cite{LSS} and \cite{GT}. There, it was proven
instability for the Brown and Ravenhall model in the free picture. In
\cite{GT} it is even deduced from this result that instability also
holds in QED context, i.e. for the Brown and Ravenhall model in the
free picture, coupled to the quantized radiation field with or without
cutoff.

In our case, working in the Furry picture and imposing both electronic
and photonic cutoffs prevent from instability.

Our methods follow those of \cite{ref2, ref3, ref4, ref5, GJ}, in
which the spectral theory of the spin-boson and Pauli-Fierz
Hamiltonians is studied.

The infrared conditions on the cutoff functions with respect to
the momenta of photons are stronger than those appearing in
\cite{ref3} and \cite{ref11}.

Some of the results obtained in this paper have been announced in
\cite{ref1} and \cite{BDG}. The case where we are only dealing with
the Wick ordering corrected Coulomb interaction of charge density with
itself can be found in \cite{bdg2}.

\medskip

\noindent{\bf Acknowledgements:} The first author was supported in
part by Minist\`ere de l'Edu\-ca\-tion Nationale, de la Recherche et
de la Technologie via ACI Blanche project. This work was supported in
part by EU TMR - Grant 96-2001. One of us (J.-C. G.)  wishes to thank
Volker Bach for helpful discussions and Heinz Siedentop for a valuable
remark. We wish to thank Beno\^\i t Gr\'ebert and Fran\c{c}ois
Nicoleau for many helpful remarks

%%%%%%%%%%%%%%%%%%%%%%%%%%%%%%%%%%%%%%%%%%%%%%%%%%%%%%%%%%%%%%%%%
%%%%%%%%%%%%%%  QUANTIZATION OF DIRAC-COULOMB FIELD %%%%%%%%%%%%%
%%%%%%%%%%%%%%%%%%%%%%%%%%%%%%%%%%%%%%%%%%%%%%%%%%%%%%%%%%%%%%%%%
\section{The quantization of the Dirac-Coulomb field}\label{S2}

The Dirac electron of mass $m_0$ in a Coulomb potential is described by the
following Hamiltonian:
\begin{equation}\nonumber
H_D := H_D(e) = c\boldsymbol{\alpha}
\cdot \frac{1}{i}\nabla + \beta m_0 c^2 - \frac{e^2 Z}{|x|},
\end{equation}
acting in the Hilbert space $L^2(\er^3;\C^4)$ with domain $\D(H_D) =
H^1(\er^3;\C^4)$, the Sobolev space of order 1. Here $\hbar =1$. We refer to
\cite{ref8} for a discussion of the Dirac operator (see also \cite{ref9}).

Here $e$ is the negative charge of the electron, $-Ze$ is the positive charge
of the nucleus and $\alpha = (\alpha_1,\alpha_2,\alpha_3)$,  $\beta$ are the
Dirac matrices in the standard form: $$ \beta = \left(\begin{array}{cc} I & 0\\
0 &-I
\end{array}\right), \qquad
\alpha_i = \left(\begin{array}{cc} 0 & \sigma_i\\ \sigma_i & 0
\end{array}\right), \qquad
i = 1, 2, 3, $$ where $\sigma_i$ are the usual Pauli matrices.

\pn It is well known that $H_D(e)$ is self-adjoint for $Z \leq
118$ (\cite[Theorem 4.4]{ref8}). The eigenstates of the
Dirac-Coulomb operator $H_D$ are labelled by the angular momentum
quantum numbers $j = \frac{1}{2}, \frac{3}{2}, \frac{5}{2},
\ldots,\ m_j = -j, -j+1, \ldots, j-1, j,$ by the spin-orbit
quantum number $k_j = \pm ( j+ \frac{1}{2}) = \pm 1, \pm 2,
\ldots$ and by the quantum number $n = 0,1,2,\ldots$, counting the
non-degenerate eigenvalues of the Dirac radial operator associated
with $k_j$. From now on we set $\gamma = (j,m_j,k_j)$.

Let $\psi_{\gamma,n}$ denote the eigenstates of the Hamiltonian $H_D$. We have
$$ H_D(e) \psi_{\gamma,n} = E_{\gamma,n}\psi_{\gamma,n}, $$ with
\begin{equation}\nonumber
E_{\gamma,n} = m_0 c^2 \left( 1 + \frac{(Ze^2)^2/c^2}{(n + \sqrt{k_j^2 -
(Ze^2)^2/c^2})^2}\right)^{-1/2}
\end{equation}
Each eigenvalue is degenerate. The infimum of the discrete spectrum is an
isolated eigenvalue of multiplicity two. Together with the bound state energy
levels the continuum energy levels are given by
\begin{equation}\label{eq:3}
\pm \omega(p), \qquad  \omega(p)=(m_0^2 c^4 + p^2)^{\frac{1}{2}}, \qquad p=|{ {\bf p}}|
\end{equation}
Here {{\bf p}} is the momentum of the electron.

Let $\psi_{\gamma,\pm}(p,x)$ denote the continuum eigenstates of $H_D(e)$. We
then have
 $$
  H_D
  \psi_{\gamma,\pm}(p,x) = \pm\omega(p)\psi_{\gamma,\pm}(p,x) .
 $$
Here, $\psi_{\gamma,n}$ and $\psi_{\gamma,\pm}(p,x)$ are normalized in such a
way that
\begin{eqnarray*}
\int_{\er^3} \psi^\dagger_{\gamma,n}(x) \psi_{\gamma',n'}(x) \d ^3
x &=&\delta_{nn'}\delta_{\gamma\gamma'},\\ \int_{\er^3}
\psi^\dagger_{\gamma,\pm}(p,x) \psi_{\gamma',\pm}(p',x) \d ^3 x
&=&\delta_{\gamma\gamma'}\delta(p-p'),\\ \int_{\er^3}
\psi^\dagger_{\gamma,\pm}(p,x) \psi_{\gamma',\mp}(p',x) \d^3 x
&=&0,\\ \int_{\er^3} \psi^\dagger_{\gamma,n}(x)
\psi_{\gamma',\pm}(p,x) \d^3 x &=&0.
\end{eqnarray*}

Here $\psi_{\gamma,\pm}^\dagger(p,x)$ (resp. $\psi_{\gamma,n}^\dagger(x)$) is
the adjoint spinor of $\psi_{\gamma,\pm}(p,x)$ (resp. $\psi_{\gamma,n}(x)$).
The spinors $(\psi_{\gamma,n})_{\gamma,n}$ and $(\psi_{\gamma,\pm})_\gamma$
generate a spectral representation of $H_D(e)$.

According to the hole theory \cite{CDG, IZ, ref10, ref9, Schweber,
ref8, W}, the absence in the Dirac theory of an electron with
energy $E < 0$ and charge $e$ is equivalent to the presence of a
positron with energy $-E > 0$ and charge $-e$.

Let us split the Hilbert space ${\mathfrak H} = L^2(\er^3 ; \C^4)$
into
 $$
  {\mathfrak H}_{c^-} = P_{(-\infty,0]}(H_D){\mathfrak H},
  \qquad {\mathfrak H}_d = P_{(0,m_0c^2)}(H_D){\mathfrak H},
  \qquad {\mathfrak H}_{c^+} = P_ {[m_0 c^2,
  +\infty)}(H_D){\mathfrak H}.
 $$
Here $P_I(H_D)$ denotes the spectral projection of $H_D$
corresponding to the interval $I$. Since
$(\psi_{\gamma,n})_{\gamma,n}$ is a basis in ${\mathfrak H}_d$,
every $\psi \in {\mathfrak H}_d$ can be written as
 $$
  \psi(x) = {\rm L.i.m}\ \sum_{\gamma,n} b_{\gamma,n}\,
  \psi_{n,\gamma}(x) ,
 $$
with
 $$
  \sum_{\gamma,n} |b_{\gamma, n}|^2 < \infty.
 $$
Hence we can identify ${\mathfrak H}_d$ with the Hilbert space $\tilde
{\mathfrak H}_d = \oplus_\gamma\, F_\gamma$, where $F_\gamma$ is a closed
subspace of $L^2(\er_+)$. Precisely, there exists a unitary operator $U_d$ from
${\mathfrak H}_d$ onto $\tilde {\mathfrak H}_d$ given by $$ U_d \psi = \left(
\sum_n \, b_{\gamma,n}\, \1_{[n,n+1)}(p) \right)_\gamma, \qquad \psi \in
{\mathfrak H}_d $$ where $p \in \er_+$.

Similarly, ${\mathfrak H}_{c\pm}$ can be identified with $\tilde {\mathfrak
H}_{c\pm} := \oplus_\gamma L^2(\er_+)$ by using the unitary operators
$U_{c\pm}$ : $$ U_{c\pm}\psi = \left( {\rm L.i.m}\, \int\,
\psi_{\gamma,\pm}^\dagger(p,x)\, \psi(x) \d x \right)_\gamma, $$ where $\psi
\in {\mathfrak H}_{c\pm}$.

\subsection{The Fock space for electrons and positrons}
$\ $

Let ${\mathfrak G} $ be any separable Hilbert space. Let $\otimes_a^n
{\mathfrak G}$ denotes the antisymmetric $n$-th tensor power of ${\mathfrak G}$
appropriate to Fermi-Dirac statistics. We define the Fermi-Fock space over
${\mathfrak G}$, denoted by ${\mathfrak F}_a({\mathfrak G})$, to be the direct
sum
 $$
  {\mathfrak F}_a({\mathfrak G}) = {
  \displaystyle\mathop \oplus_{n=0}^\infty} \, \otimes_a^n {\mathfrak G} ,
 $$
where $\otimes_a^0{\mathfrak G} := \C$. The state $\Omega$ will denote the
vacuum vector, i.e., the vector
 $(1,0,0,\cdots)$.

Let ${\mathfrak F}_{a,d}$, ${\mathfrak F}_{a,+}$ and ${\mathfrak F}_{a,-}$ be
the Fermi-Fock spaces over $\tilde {\mathfrak H}_d$, $\tilde {\mathfrak
H}_{c+}$ and $\tilde {\mathfrak H}_{c-}$ respectively. $\Omega_d$, $\Omega_+$
and $\Omega_-$ denote the associated vacua.

In the Furry picture, the Fermi-Fock space for electrons and positrons, denoted
by ${\mathfrak F}_D$, is then the following Hilbert space
\begin{eqnarray}
{\mathfrak F}_D = {\mathfrak F}_{a,d} \otimes {\mathfrak F}_{a,+} \otimes
{\mathfrak F}_{a,-}
\end{eqnarray}
The vector $\Omega_d \otimes \Omega_+ \otimes \Omega_-$ is the vacuum of
electrons and positrons.

One has $$ {\mathfrak F}_D = \displaystyle \mathop \oplus_{q,r,s=0}^\infty
{\mathfrak F}_a^{(q,r,s)} $$ where ${\mathfrak F}_d^{(q,r,s)} = (\otimes_a^q
\tilde {\mathfrak H}_d) \otimes (\otimes_a^r \tilde {\mathfrak H}_{c+}) \otimes
(\otimes_a^s \tilde {\mathfrak H}_{c-})$.

Let us remark that ${\mathfrak F}_D$ is unitarily equivalent to ${\mathfrak
F}_a(\tilde {\mathfrak H}_d \oplus \tilde {\mathfrak H}_{c+} \oplus \tilde
{\mathfrak H}_{c-})$. (See \cite{D} and \cite{DG}).

\subsection{Creation and annihilation operators}
For every $\varphi \in {\mathfrak H}$ we define in ${\mathfrak F}_a({\mathfrak
H})$ the annihilation operator, denoted by $b(\varphi)$, which maps
$\otimes_a^{n+1} {\mathfrak H}$ into $\otimes_a^n {\mathfrak H}$ :
\begin{eqnarray*}
&& b(\varphi)\, (A_{n+1}(\varphi_1 \otimes \ldots \otimes \varphi_{n+1}))\\
\noalign{\vskip 8pt} && \qquad = \frac{\sqrt{n+1}}{(n+1)!} \sum_\sigma
sgn(\sigma)\, (\varphi, \varphi_{\sigma(1)})\, \varphi_{\sigma(2)}\otimes
\ldots \otimes \varphi_{\sigma(n+1)}
\end{eqnarray*}
where $\varphi_i \in {\mathfrak H}$.

The creation  operator, denoted by $b^*(\varphi)$, is the adjoint of
$b(\varphi)$. The operators $b^*(\varphi)$ and $b(\varphi)$ are bounded in
${\mathfrak F}_a({\mathfrak H})$ and $\|b(\varphi)\| = \|b^*(\varphi)\| =
\|\varphi\|$.

We now define the annihilation and creation operators in ${\mathfrak F}_D$.

\subsubsection{Bound states}\label{S221}
For every $(\gamma, n)$, we define in ${\mathfrak F}_D$ the annihilation
operator, denoted by $b_{\gamma,n}$, which maps ${\mathfrak F}^{(q+1,r,s)}$
into ${\mathfrak F}^{(q,r,s)}$ :
\begin{eqnarray*}
b_{\gamma,n}\left( A_{q+1}(f_1\otimes \ldots \otimes f_{q+1}) \otimes A_r(g_1
\otimes \ldots \otimes g_r) \otimes A_s(h_1 \otimes \ldots \otimes h_s)
\right)&&\\ \noalign{\vskip 8pt} = \left[ b(U_d\, \psi_{\gamma,n}) A_{q+1}(f_1
\otimes \ldots \otimes f_{r+1}) \right] \otimes A_r(g_1 \otimes \ldots \otimes
g_r) \otimes A_s(h_1 \otimes \ldots \otimes h_s)&&
\end{eqnarray*}
where $f_i \in \tilde {\mathfrak H}_d$, $g_j \in \tilde {\mathfrak H}_{c+}$ and
$h_k\in \tilde {\mathfrak H}_{c^-}$.

The creation operator $b_{\gamma,n}^*$ is the adjoint of $b_{\gamma,n}$.

$b_{\gamma,n}^*$ and $b_{\gamma,n}$ are bounded operators in ${\mathfrak F}_D$.

\subsubsection{Electrons in the continuum}
For every $g\in \tilde {\mathfrak H}_{c+}$, we define in ${\mathfrak F}_D$ the
annihilation operator, denoted by $b_+(g)$, which maps ${\mathfrak
F}^{(q,r+1,s)}$ into ${\mathfrak F}^{(q,r,s)}$ as follows
\begin{eqnarray*}
 b_+(g) \left( A_q(f_1 \otimes \ldots \otimes f_q) \otimes
 A_{r+1}(g_1 \otimes \ldots \otimes g_{r+1}) \otimes A_s(h_1 \otimes \ldots
 \otimes h_s) \right)&&\\
 \noalign{\vskip 8pt} = A_q(f_1 \otimes \ldots \otimes
 f_q) \otimes \left[ (-1)^q \, b(g) A_{r+1}
 (g_1 \otimes \ldots \otimes g_{r+1})
 \right] \otimes A_s(h_1 \otimes \ldots \otimes h_s)&&
\end{eqnarray*}
The creation operator $b_+^*(g)$ is the adjoint of $b_+(g)$. $b_+^*(g)$ and
$b_+(g)$ are bounded operators in ${\mathfrak F}_D$.

We set, for every $\psi \in \tilde {\mathfrak H}_{c+}$,
\begin{eqnarray*}
b_{\gamma,+}(\psi)   &=& b_+  (P_\gamma^+ \psi)\\ \noalign{\vskip 8pt}
b_{\gamma,+}^*(\psi)&=& b_+^*(P_\gamma^+ \psi)
\end{eqnarray*}
where $P_\gamma^+$ is the projection of $\tilde {\mathfrak H}_{c+}$ onto the
$\gamma$-th component.

\subsubsection{Positrons}
For every $h \in \tilde {\mathfrak H}_{c-}$, we define in ${\mathfrak F}_D$ the
annihilation operator, denoted by $b_-(h)$, which maps ${\mathfrak
F}^{(q,r,s+1)}$ into ${\mathfrak F}^{(q,r,s)}$ :
\begin{eqnarray*}
b_-(h) (A_q(f_1 \otimes \ldots \otimes f_q) \otimes A_r(g_1 \otimes
\ldots\otimes g_r) \otimes A_{s+1}(h_1 \otimes \ldots \otimes h_{s+1}))&&\\
\noalign{\vskip 8pt} = A_q(h_1 \otimes \ldots \otimes f_q) \otimes A_r(g_1
\otimes \ldots \otimes g_r) \otimes [(-1)^{q+r} b(h) A_{s+1}(h_1 \otimes \ldots
\otimes h_{s+1})]&&
\end{eqnarray*}
The creation operator $b_-^*(h)$ is the adjoint of $b_-(h)$. $b_-^*(h)$ and
$b_-(h)$ are bounded operators in ${\mathfrak F}_D$.

We set for every $\psi \in \tilde {\mathfrak H}_{c-}$,
\begin{eqnarray*}
b_{\gamma,-}(\psi)   &=& b_{-}  (P_\gamma^- \psi)\\ \noalign{\vskip 8pt}
b_{\gamma,-}^*(\psi) &=& b_-^*(P_\gamma^- \psi)
\end{eqnarray*}
where $P_\gamma^-$ is the projection of $\tilde {\mathfrak H}_{c-}$ onto the
$\gamma$-th component.

A simple computation shows that the following anti-commutation relations hold
\begin{eqnarray*}
&& \{b_{\gamma,n}^*, b_{\beta,m}\} = \delta_{\gamma,\beta}\, \delta_{n,m}\\
\noalign{\vskip 8pt} && \{b_{\gamma,n}^*, b_{\beta,m}^*\}=0 = \{b_{\gamma,n},
b_{\beta,m}\}\\ \noalign{\vskip 8pt} && \{b_{\gamma,+}(\psi_1),
b_{\beta,+}^*(\psi_2)\} = \delta_{\gamma,\beta}(P_\gamma^+ \psi_1, P_\gamma^+
\psi_2)_{L^2(\er_+)}, \qquad \psi_i \in \tilde {\mathfrak H}_{c+}\\
\noalign{\vskip 8pt} && \{b_{\gamma,-}(\psi_1), b_{\beta,-}^*(\psi_2)\} =
\delta_{\gamma,\beta}(P_\gamma^- \psi_1, P_\gamma^- \psi_2)_{L^2(\er_+)},
\qquad \psi_i \in \tilde {\mathfrak H}_{c-}
\end{eqnarray*}
Furthermore
\begin{equation}\nonumber
\{ b_{\gamma,\pm}^\#(\psi), b_{\beta,n}^\#\} = 0.
\end{equation}
Here $b^\#$ is $b$ or $b^*$.
\begin{eqnarray*}
\{b_{\gamma,+}(\psi_1), b_{\beta,-}(\psi_2)\} = \{b_{\gamma,+}(\psi_1),
b_{\beta,-}^*(\psi_2)\} = 0,\\ \noalign{\vskip 8pt} \{b_{\gamma,+}^*(\psi_1),
b_{\beta,-}(\psi_2)\} = \{b_{\gamma,+}^*(\psi_1), b_{\beta,-}^*(\psi_2)\} = 0,
\end{eqnarray*}
where $\psi_1 \in \tilde {\mathfrak H}_{c+}$ and $\psi_2 \in
\tilde {\mathfrak H}_{c-}$. One should remark that, in contrast to
\cite{ref8}, the charge conjugation operator is not included in
the definition of the annihilation and creation operators for the
positrons.

Our definition is close to the hole theory and is the one occurring in
many text books in Quantum Field Theory as in \cite{IZ}, \cite{ref10},
\cite{Schweber} and \cite{W}. The other method for the quantization of
the Dirac field is close to the so-called symmetric theory of
charge. Both approaches are described in \cite{ref10}.

As in \cite[chapter X]{RS}, we introduce
operator-valued distributions $b_{\gamma,\pm}(p)$ and $b_{\gamma,\pm}^*(p)$
such that we write
\begin{eqnarray*} b_{\gamma,\pm}(\psi)   &=& \int_{\er^+}\, \d p\
b_{\gamma,\pm}(p)\,  \overline{(P_\gamma^\pm \psi)\, (p)}\\ \noalign{\vskip
8pt} b_{\gamma,\pm}^*(\psi) &=& \int_{\er^+}\, \d p\ b_{\gamma,\pm}^*(p)\,
(P_\gamma^\pm \psi)\, (p) \end{eqnarray*} where $\psi \in \tilde {\mathfrak
H}_{c+} \oplus \tilde {\mathfrak H}_{c-}$.

We now give a representation of $b_{\gamma,\pm}(p)$ and $b_{\gamma,\pm}^*(p)$.
Let $\D_D$ denote the set of smooth vectors $\Phi \in {\mathfrak F}_D$
 for which $\Phi^{(q,r,s)}$ has a compact  support and $\Phi^{(q,r,s)} =
0$ for all but finitely
 many $(q,r,s)$.

For every $(p,\gamma)$, $b_{\gamma,+}(p)$ maps ${\mathfrak F}_a^{(q,r+1,s)}
\cap \D_D$ into
 ${\mathfrak F}_a^{(q,r,s)} \cap \D_D$ and we have
\begin{eqnarray*}
 (b_{\gamma,+}(p)\Phi)^{(q,r,s)}(p_1,\gamma_1,\ldots, p_q,\gamma_q ;
 p_1,\gamma_1,\ldots, p_r,\gamma_r ; p_1,\gamma_1,\ldots, p_s,\gamma_s)&&\\
  = \sqrt{r+1}(-1)^q \Phi^{(q,r+1,s)} (p_1,\gamma_1,\ldots,
 p_q,\gamma_q;p,\gamma, p_1,\gamma_1,\ldots, p_r,\gamma_r ;p_1,\gamma_1,\ldots,
 p_s,\gamma_s)&&
\end{eqnarray*}
$b_{\gamma,+}^*(p)$ is then given by:
\begin{eqnarray*}
&&(b_{\gamma,+}^*(p)\Phi)^{(q,r+1,s)}(p_1,\gamma_1,\ldots, p_q,\gamma_q ;
p_1,\gamma_1,\ldots, p_{r+1},\gamma_{r+1} ; p_1,\gamma_1,\ldots,
p_s,\gamma_s)=\\
 &&\frac{(-1)^q}{\sqrt{r+1}}\sum_{i=1}^{r+1}(-1)^{i+1}
\delta_{\gamma_i \gamma}\delta(p-p_i)\\ &&\Phi^{(q,r,s)}(p_1,\gamma_1,\ldots,
p_q,\gamma_q; p_1,\gamma_1,\ldots, \widehat{p_i,\gamma_i},\ldots,
p_{r+1},\gamma_{r+1} ; p_1,\gamma_1,\ldots, p_s,\gamma_s)
\end{eqnarray*}
where $\widehat\cdot$ denotes that the i-th variable has to be
omitted.

Similarly $b_{\gamma,-}(p)$ maps ${\mathfrak F}_a^{(q,r,s+1)} \cap
\D_D$ into ${\mathfrak F}_a^{(q,r,s)} \cap \D_D$ such that
\begin{equation}\nonumber
\begin{split}
 \lefteqn{(b_{\gamma,-}(p)\Phi)^{(q,r,s)}(p_1,\gamma_1,\ldots, p_q,\gamma_q
 ; p_1,\gamma_1,\ldots, p_r,\gamma_r ; p_1,\gamma_1,\ldots,
 p_s,\gamma_s)=} &\\
 & \sqrt{s+1}(-1)^{q+r} \Phi^{(q,r,s+1)}
 (p_1,\gamma_1,\ldots, p_q,\gamma_q ; p_1,\gamma_1,\ldots,
 p_r,\gamma_r ; p, \gamma, p_1,\gamma_1,\ldots, p_s,\gamma_s)
\end{split}
\end{equation}
$b_{\gamma,-}^*(p)$ is then given by
\begin{eqnarray*}
&& (b_{\gamma,-}^*(p)\Phi)^{(q,r,s+1)} (p_1,\gamma_1,\ldots, p_q,\gamma_q ;
p_1,\gamma_1,\ldots, p_r,\gamma_r ; p_1,\gamma_1,\ldots, p_{s+1},\gamma_{s+1})=
\\ &&\frac{1}{\sqrt{s+1}}(-1)^{q+r} \sum_{i=1}^{s+1}(-1)^{i+1}
\delta_{\gamma,\gamma_i}\delta(p-p_i) \\ &&\Phi^{(q,r,s)}(p_1,\gamma_1,\ldots,
p_q,\gamma_q; p_1,\gamma_1,\ldots, p_r,\gamma_r ; p_1,\gamma_1,\ldots,
\widehat{p_i,\gamma_i},\ldots, p_{s+1},\gamma_{s+1})
\end{eqnarray*}
Let us recall that $\Phi^{(q,r,s)}$ is antisymmetric in the bound states, the
electron in the continuum and the positron variables separately. We have
\begin{equation}\nonumber
\{b_{\gamma,+}(p), b_{\gamma',+}^*(p')\} = \{b_{\gamma,-}(p),
b_{\gamma',-}^*(p')\} = \delta_{\gamma,\gamma'}\delta(p-p')
\end{equation}
Any other anti-commutation relation is equal to zero. In particular we have
\begin{equation}\nonumber
\{b_{\gamma,\pm}(p), b_{\gamma',n}^\#\} = \{b_{\gamma,\pm}^*(p),
b_{\gamma',n}^\#\} = 0
\end{equation}
where $b^\#$ is $b$ or $b^*$.

\subsubsection{The Hamiltonian for the quantized Dirac-Coulomb field}
The quantization of the Dirac-Coulomb Hamiltonian $H_D$, denoted by
$d\Gamma(H_D)$, is now given by
\begin{eqnarray*}
 d\Gamma(H_D) &=& \sum_{\gamma,n} E_{\gamma,n}\, b_{\gamma,n}^*\, b_{\gamma,n} +
 \sum_\gamma \int \d p\ \omega(p)\, b_{\gamma,+}^*(p)\, b_{\gamma,+}(p)
 \nonumber\\
 \noalign{\vskip 8pt} \qquad && + \sum_\gamma \int \d p\ \omega(p)\,
 b_{\gamma,-}^*(p)\, b_{\gamma,-}(p) ,
\end{eqnarray*}
with $\omega(p)$ given in \eqref{eq:3}. The operator $d\Gamma(H_D)$ is the
Hamiltonian of the quantized Dirac-Coulomb field. It is well defined on the
dense subset $\D_D$ and it is essentially self-adjoint on $\D_D$. The
self-adjoint extension will be also denoted by $d\Gamma(H_D)$ with domain
$\D(d\Gamma(H_D))$. Moreover the operator number of electrons and positrons,
denoted by $N_D$, is given by
 $$
  N_D = \sum_{\gamma, n} b_{\gamma,n}^*
  b_{\gamma,n} + \sum_\gamma \int \d p\
  b_{\gamma,+}^*(p) b_{\gamma,+}(p) +
  \sum_\gamma \int \d p\ b_{\gamma,-}^*(p) b_{\gamma,-}(p) .
 $$
The operator $N_D$ is essentially self-adjoint on $\D_D$. The self-adjoint
extension will be also denoted by $N_D$ with domain $\D(N_D)$.\\ Let $\Delta$
denote the set of all eigenvalues of the operator
 $$
  d\Gamma(H_D)_d := \sum_{\gamma,n} E_{\gamma,n}
  b_{\gamma,n}^* b_{\gamma,n}
 $$
Each $E_{\gamma, n}$ is in $\Delta$ and we have
\begin{equation}\nonumber
0 < E_0 = \inf_{\gamma,n} E_{\gamma,n}, \,\,\,\,\, {\rm and }\,\,\, \inf_{E\in
\Delta} E=0.
\end{equation}
$0$ is the eigenvalue associated with $\Omega_D$. Furthermore
 $$
  \sigma(d\Gamma(H_D)) = \Delta \cup [m_0 c^2, \infty) .
 $$
The set $[m_0 c^2, \infty)$ is the absolutely continuous spectrum of
$d\Gamma(H_D)$. Let us remark that, for $E \in \Delta$ with $E > m_0 c^2$, $E$
is an eigenvalue embedded in the continuous spectrum.

\subsection{The Fock space for transversal photons}
The one photon Hilbert space is given by $L^2(\er^3,\C^2)$. For every $f$ in
$L^2(\er^3, \C^2)$, we shall write $f(k,\mu)$ where $k \in \er^3$ is the
momentum variable of the photon and $\mu = 1,2$ is its polarization index
associated with two given independent real transversal polarizations
$\varepsilon_\mu(k)$ of the photon in the Coulomb gauge such that
$\varepsilon_\mu(k).\varepsilon_{\mu'}(k) = \delta_{\mu \mu'}$ and
$\varepsilon_\mu(k).k=0$.

Let ${\mathfrak F}_{ph}$ denote the Fock space for transversal photons : $$
{\mathfrak F}_{ph} = \oplus_{n=0}^\infty L^2(\er^3,\C^2)^{\otimes_s^n} $$ where
$L^2(\er^3,\C^2)^{\otimes_s^0} = \C$. Here $L^2(\er^3,\C^2)^{\otimes_s^n}$ is
the symmetric $n$-tensor power of $L^2(\er^3,\C^2)$ appropriate for
Bose-Einstein statistics.

For $f\in L^2(\er^3,\C^2)$, the annihilation and creation operators, denoted by
$a(f)$ and $a^*(f)$ respectively, are now given by:
\begin{eqnarray*}
a(f)   &=& \sum_{\mu =1,2} a_\mu  (f(\cdot,\mu))\\ \noalign{\vskip 8pt} a^*(f)
&=& \sum_{\mu =1,2} a_\mu^*(f(\cdot,\mu))
\end{eqnarray*}
where
\begin{eqnarray*}
&&(a_\mu(f(\cdot,\mu))\Psi)^{(n)} (k_1,\mu_1,k_2,\mu_2,\ldots, k_n,\mu_n)=\\
\noalign{\vskip 8pt} &&\qquad \sqrt{n+1} \int \d^3k\, \overline{f(k,\mu)}\,
\Psi^{(n+1)} (k,\mu, k_1,\mu_1, k_2,\mu_2,\ldots, k_n,\mu_n)
\end{eqnarray*}
\begin{eqnarray*}
&&(a_\mu^*(f(\cdot,\mu))\Psi)^{(n)} (k_1,\mu_1,\ldots, k_n,\mu_n)=\\
\noalign{\vskip 8pt} &&\qquad \frac{1}{\sqrt{n}} \sum_{i=1}^{n}\,
f(k_i,\mu_i)\, \Psi^{(n-1)} (k_1,\mu_1, \ldots, \widehat{k_i,\mu_i},\ldots,
k_n,\mu_n)
\end{eqnarray*}
where $\widehat{\cdot}$ denotes that the $i$-th variable has to be omitted.
Note that $a_\mu^*(f(\cdot,\mu))$ and $a_\mu(f(\cdot,\mu))$ are linear and
anti-linear with respect to $f$ respectively, so that we can introduce operator
valued distributions, i.e, fields $a_\mu(k)$ and $a_\mu^*(k)$ such that $$
a_\mu(f(\cdot,\mu)) = \int \d^3 k\ \overline{f(k,\mu)}\, a_\mu (k) $$ and $$
a_\mu^*(f(\cdot,\mu)) = \int \d^3 k\ f(k,\mu)\, a_\mu^* (k) $$ where
$f(\cdot,\mu) \in L^2(\er^3)$.

Let $\D_{ph}$ denote the set of smooth vectors $\Psi \in {\mathfrak F}_{ph}$
for which $\Psi^{(n)}$ has a compact support and
 $\Psi^{(n)} = 0$
for all but finitely many $n$. Then, for any $\Psi \in \D_{ph},$ the action of
$a_\mu(k)$ and $a_\mu^*(k)$ is given by
\begin{eqnarray*}
&&(a_\mu(k)\Psi)^{(n)} (k_1,\mu_1,k_2,\mu_2,\ldots, k_n,\mu_n)\\
\noalign{\vskip 8pt} &&\qquad =\sqrt{n+1} \Psi^{(n+1)} (k,\mu,
k_1,\mu_1,\ldots, k_n,\mu_n)
\end{eqnarray*}
\begin{eqnarray*}
&&(a_\mu^*(k)\Psi)^{(n+1)} (k_1,\mu_1;k_2,\mu_2,\ldots, k_{n+1},\mu_{n+1})\\
\noalign{\vskip 8pt} &&\qquad = \frac{1}{\sqrt{n+1}} \sum_{i=1}^{n+1}\,
\delta_{\mu_i,\mu}\, \delta(k_i-k)\, \Psi^{(n)} (k_1,\mu_1, \ldots,
\widehat{k_i,\mu_i},\ldots, k_{n+1},\mu_{n+1})
\end{eqnarray*}
We have the canonical commutation relations
\begin{equation}\nonumber
[a_\mu(k), a_{\mu'}^*(k')] = \delta_{\mu\mu'} \delta(k-k')
\end{equation}
Any other commutation relation is equal to zero.

The Hamiltonian of the quantized electromagnetic field, denoted by $H_{ph}$, is
$$ H_{ph} = \sum_{\mu =1,2} \int \d^3 k\ \omega(k)\, a_\mu^*(k)\, a_\mu(k) $$
where $\omega(k) = c|k|$. $H_{ph}$ is essentially self-adjoint on $\D_{ph}$.

The state $(1,0,\ldots) \in {\mathfrak F}_{ph}$ is the vacuum of photons and
will be denoted $\Omega_{ph}$.

The spectrum of $H_{ph}$ consists of an absolutely continuous spectrum covering
$\lbrack 0, +\infty)$ and a simple eigenvalue, equal to zero, whose the
corresponding eigenvector is  the vacuum state  $\Omega \in {\mathfrak
F}_{ph}$.
\subsection{The total Hamiltonian and the main results}\label{S2.4}
The Fock space for electrons,
positrons and photons is the following Hilbert  space:
\begin{equation}\nonumber
{\mathfrak F} = {\mathfrak F}_D \otimes {\mathfrak F}_{ph}.
\end{equation}
$\Omega = \Omega_D \otimes \Omega_{ph}$ is the vacuum of ${\mathfrak F}$.

The free Hamiltonian for electrons, positrons and photons, denoted by $H_0$, is
the following operator in ${\mathfrak F}$ :
\begin{equation}\nonumber
H_0 = d\Gamma(H_D) \otimes \1_{ph} + \1_D \otimes H_{ph}
\end{equation}
$H_0$ is a self-adjoint operator with  domain
 $$
  \D\left(d\Gamma(H_D) \otimes \1_{ph}\right)
  \cap \D\left(\1_D \otimes H_{ph}\right) .
 $$
The operator $H_0$ has the same point spectrum as $d\Gamma(H_D)$ and its
continuous spectrum covers the half axis $[0,\infty)$. Hence the point spectrum
of $d\Gamma(H_D)$ is embedded in the continuous spectrum of $H_0$. The
eigenfunctions of $H_0$ corresponding to the eigenvalues $E$ have the form
$\varphi \otimes \Omega_{ph}$, where $\varphi$ are the eigenfunctions of
$d\Gamma(H_D)$ corresponding to the eigenvalues $E$.

Let us now describe the interaction between electrons, positrons and photons in
Coulomb gauge that we consider.

Let us first recall the physical interaction in QED with the
Coulomb gauge.

The quantized Dirac-Coulomb field is given by
\begin{eqnarray*}
 \psi(x) &=& \sum_{\gamma,n} b_{\gamma,n} \psi_{\gamma,n}(x)\\ \noalign{\vskip
 8pt} && + \sum_\gamma \int_{\er^+} \d p\ b_{\gamma,+}(p) \psi_{\gamma,+}(p,x) +
 \sum_\gamma \int_{\er^+} \d p\ b_{\gamma,-}^*(p) \psi_{\gamma,-}(p,x)
\end{eqnarray*}
together with
\begin{eqnarray*}
 \psi^\dagger(x) &=& \sum_{\gamma,n} b_{\gamma,n}^* \psi_{\gamma,n}^\dagger(x)+
 \sum_\gamma \int_{\er^+} \d p\ b_{\gamma,+}^*(p) \psi_{\gamma,n}^\dagger(p,x)\\
 \noalign{\vskip 8pt} && + \sum_\gamma \int_{\er^+} \d p\ b_{\gamma,-}(p)
 \psi_{\gamma,-}^\dagger(p,x)
\end{eqnarray*}
for the quantized adjoint spinor.

Here $b_{\gamma,n}$, $b_{\gamma,n}^*$, $b_{\gamma,\pm}(p)$ and
$b_{\gamma,\pm}^*(p)$ are the annihilation and creation operators defined
above.

The density of charge is then given by $$ \rho(x) = e :\psi^\dagger(x)\,
\psi(x): $$ and the density of current by $$ j(x) = ec :\psi^\dagger(x)\,
\alpha\, \psi(x): $$ Here $:\ :$ is the normal ordering and $\boldsymbol{\alpha} =
(\alpha_1,\alpha_2,\alpha_3)$.

The interaction of electrons and positrons with photons is given
by two terms. The first one is
\begin{equation}\label{interaction1}
  \frac{e^2}{2} \int_{\er^3\times \er^3} \d ^3 x\, \d^3 x'\
  \frac{\rho(x)\, \rho(x')}{|x-x'|} ,
\end{equation}
and describes the Coulomb interaction of densities of charges. The
second one,
\begin{equation}\nonumber
  -e \int_{\er^3} \d^3 x\ j(x)\cdot A(x) ,
\end{equation}
describes the interaction between the current and the transversal photons. Here
$A(x)$ is the quantized electromagnetic field in Coulomb gauge
 $$
  A(x) = \sum_{\mu =1,2} \left( \frac{\hbar}
  {2\varepsilon_0(2\pi)^3}\right)^{1/2} \int
  \frac{\d^3 k}{\sqrt{2\omega_{ph}}(k)}
  \left( \varepsilon_\mu(k) e^{ik\cdot x}
  a_\mu(k) + \varepsilon_\mu(k) e^{-ik\cdot x} a_\mu^*(k)\right)
 $$
(See \cite{CDG} and \cite{ref6}), $a_\mu^*(k)$ and $a_\mu(k)$ are the creation and annihilation
operators on ${\mathfrak F}_{ph}$.

Then, at a formal level, the interaction terms can be expressed in
terms of the annihilation and creation operators $b_{\gamma, n}$,
$b_{\gamma,n}^*$, $b_{\gamma, \pm}(p)$, $b_{\gamma, \pm}^* (p)$,
$a_\mu(k)$ and $a^*_\mu(k)$. Note that for the interaction between the
current and the transversal photons, the products of the $b$'s and
$b^*$'s must be Wick normal ordered. Furthermore, the Coulomb
interaction of densities of charges \eqref{interaction1} as it stands,
is not Wick normal ordered. For both physical and mathematical reasons
(see \cite{RS} and \cite{W}) it is more convenient to rewrite it as a
sum of Wick normal ordered terms by using the anti-commutation
relations.

It is a known fact that we have to introduce several cutoff functions
in the Dirac-Coulomb field and the electromagnetic vector potential in
order to get a well defined total Hamiltonian in the Fock space (see
\cite{LS}).

Thus the interaction between the electrons, positrons and photons
consists of two terms. The first term in the interaction, denoted by
$H_I^{(1)}$, is given by
\begin{eqnarray}\nonumber
\lefteqn{\quad\ \  H_I ^{(1)} =} &  & \\ & & \sum_{\gamma,\gamma',n,\ell}
\sum_{\mu =1,2} \int \d^3 k\, \bigg(G_{d,\gamma,\gamma',n,\ell}^\mu(k)
b_{\gamma,n}^* b_{\gamma',\ell} a^*_\mu(k) + {\rm h.c.}\bigg)\nonumber\\ & & +
\sum_{\epsilon=+,-}\sum_{\gamma,\gamma',n} \sum_{\mu =1,2} \int \d^3 k\,\d p\,
\bigg(G_{d,\epsilon,\gamma,\gamma',n}^\mu(p;k) \Big(b_{\gamma,n}^*
b_{\gamma',\epsilon}(p)\nonumber \\ & &  \hskip 4.6cm +
b_{\gamma,\epsilon}^*(p)b_{\gamma',n} \Big)a^*_\mu(k) + {\rm
h.c.}\bigg)\nonumber\\ & & + \sum_{\gamma,\gamma'} \sum_{\mu =1,2} \int \d^3
k\,\d p\,\d p' \bigg(G_{+,-,\gamma,\gamma'}^\mu(p,p';k)  \Big(b_{\gamma,+}^*(p)
b_{\gamma',-}^*(p') \nonumber \\ & & \hskip 3.9cm +
b_{\gamma,-}(p)b_{\gamma',+}(p') \Big) a^*_\mu(k) + {\rm h.c.}\bigg)\nonumber\\
& & + \sum_{\epsilon=+,-}\sum_{\gamma,\gamma'} \sum_{\mu =1,2} \int \d^3 k\,\d
p\,\d p' \bigg(G_{\epsilon,\epsilon,\gamma,\gamma'}^\mu(p,p';k)
b_{\gamma,\epsilon}^*(p) b_{\gamma',\epsilon}(p') a^*_\mu(k) + {\rm
h.c.}\bigg)\nonumber
\end{eqnarray}
For the second term let us introduce some notations. In the case
of electrons, $\xi$ will be equal to $(\gamma,p)$ and $(\gamma,n)$
with $\int \d\xi=\sum_\gamma \int \d p + \sum_{\gamma,n}$. In the
case of positrons, $\xi$ will be equal to $(\gamma, p)$ with $\int
\d\xi=\sum_\gamma \int \d p$ and the $L^2$ norm of any function of
$\xi$ will be the sum of the $L^2$ norm with respect to the
continuous part of the measure and the $L^2$ norm with respect to
the discrete part of the measure. The second term of the
interaction, denoted by $H_I^{(2)}$, is an operator in
$\mathfrak{F}_D$ given by
\begin{eqnarray}\nonumber
\lefteqn{\quad H_I^{(2)} =} & & \\ & & \int \d\xi_1
\d\xi_2\d\xi_3\d\xi_4\, F^{(1)}(\xi_1,\xi_2,\xi_3,\xi_4)
b_+^*(\xi_1) b_-^*(\xi_2) b_+(\xi_3)b_-(\xi_4)\nonumber\\ & & +
\sum_{\epsilon=+,-} \int \d\xi_1 \d\xi_2\d\xi_3\d\xi_4\,
F^{(2)}_\epsilon(\xi_1,\xi_2,\xi_3,\xi_4) b_\epsilon^*(\xi_1)
b_\epsilon^*(\xi_2) b_\epsilon(\xi_3)b_\epsilon(\xi_4)\nonumber\\
& & +\sum_{\epsilon,\epsilon'=+,-,\atop \epsilon\neq\epsilon'}
\int \d\xi_1 \d\xi_2 \d\xi_3 \d\xi_4\,
\Big(F^{(3)}_{\epsilon,\epsilon'} (\xi_1,\xi_2,\xi_3,\xi_4)
b_\epsilon^*(\xi_1) b_\epsilon(\xi_2)
b_{\epsilon'}(\xi_3) b_{\epsilon}(\xi_4)\nonumber \\ & &
\hskip4.2cm - \overline{F^{(3)}_{\epsilon,\epsilon'}
(\xi_4,\xi_2,\xi_3,\xi_1)} b_\epsilon^*(\xi_1) b_\epsilon^*(\xi_2)
b_{\epsilon'}^*(\xi_3) b_{\epsilon}(\xi_4)\Big)\nonumber\\ & &
+\int
\d\xi_1 \d\xi_2 \d\xi_3 \d\xi_4\, \Big(F^{(4)}
(\xi_1,\xi_2,\xi_3,\xi_4) b_+(\xi_1) b_+(\xi_2)
b_-(\xi_3) b_-(\xi_4) \nonumber\\ & &
\hskip4.2cm + \overline{F^{(4)}
(\xi_4,\xi_2,\xi_3,\xi_1)} b_+^*(\xi_1) b_+^*(\xi_2)
b_-^*(\xi_3)b_-^*(\xi_4)\Big) \nonumber \\
& & + \sum_{\epsilon =\pm} \int \d\xi_1 \d\xi_2
F_\epsilon^{(5)}(\xi_1, \xi_2) b_\epsilon^*(\xi_1) b_\epsilon(\xi_2)
\nonumber \\
& & + \sum_{\epsilon,\epsilon'=+,-,\atop \epsilon\neq\epsilon'}
 \int \d\xi_1 \d\xi_2
 \left( F_{\epsilon, \epsilon'}^{(6)} (\xi_1, \xi_2)
b_\epsilon^*(\xi_1) b_{\epsilon'}^*(\xi_2) +
\overline{F_{\epsilon, \epsilon'}^{(6)} (\xi_2, \xi_1) }
b_{\epsilon'}(\xi_1) b_\epsilon(\xi_2)\right) \nonumber
\end{eqnarray}
where, for $\xi=(\gamma, n)$, $b^\sharp(\xi) :=
b_{\gamma,n}^\sharp$.
%%%%%%  PREVIOUS FORM: %%%%%%%%%%%%%%%%%%%%%%%%%%%%%%%%%%%%%%%%%%%%
%\begin{eqnarray}\nonumber
%\lefteqn{\quad H_I^{(2)} =} & & \\ & & \int \d\xi_1
%\d\xi_2\d\xi_3\d\xi_4\, F^{(1)}(\xi_1,\xi_2,\xi_3,\xi_4)
%b_+^*(\xi_1) b_-^*(\xi_2) b_+(\xi_3)b_-(\xi_4)\nonumber\\ & & +
%\sum_{\epsilon=+,-} \int \d\xi_1 \d\xi_2\d\xi_3\d\xi_4\,
%F^{(2)}_\epsilon(\xi_1,\xi_2,\xi_3,\xi_4) b_\epsilon^*(\xi_1)
%b_\epsilon^*(\xi_2) b_\epsilon(\xi_3)b_\epsilon(\xi_4)\nonumber\\
%& & +\sum_{\epsilon,\epsilon'=+,-,\atop \epsilon\neq\epsilon'}
%\int \d\xi_1 \d\xi_2\d\xi_3\d\xi_4\,
%\Big(F^{(3)}_{\epsilon,\epsilon'} (\xi_1,\xi_2,\xi_3,\xi_4)
%b_\epsilon^*(\xi_1) b_\epsilon(\xi_2)
%b_{\epsilon'}(\xi_3)b_{\epsilon}(\xi_4)\nonumber \\ & &
%\hskip4.2cm - \overline{F^{(3)}_{\epsilon,\epsilon'}
%(\xi_4,\xi_2,\xi_3,\xi_1)} b_\epsilon^*(\xi_1) b_\epsilon^*(\xi_2)
%b_{\epsilon'}^*(\xi_3)b_{\epsilon}(\xi_4)\Big)\nonumber\\ & &
%+\sum_{\epsilon,\epsilon'=+,-,\atop \epsilon\neq\epsilon'} \int
%\d\xi_1 \d\xi_2\d\xi_3\d\xi_4\, \Big(F^{(4)}_{\epsilon,\epsilon'}
%(\xi_1,\xi_2,\xi_3,\xi_4) b_\epsilon(\xi_1) b_\epsilon(\xi_2)
%b_{\epsilon'}(\xi_3)b_{\epsilon'}(\xi_4)\nonumber\\ & &
%\hskip4.2cm + \overline{F^{(4)}_{\epsilon,\epsilon'}
%(\xi_4,\xi_2,\xi_3,\xi_1)} b_\epsilon^*(\xi_1) b_\epsilon^*(\xi_2)
%b_{\epsilon'}^*(\xi_3)b_{\epsilon'}^*(\xi_4)\Big) .\nonumber
%\end{eqnarray}
%%%%%%%%%%%%%%%%%%%%%%%%%%%%%%%%%%%%%%%%%%%%%%%%%%%%%%%%%%%%%%%%%%%
Furthermore, we suppose that
 $$
 \begin{array}{l}
   F^{(1)}(\xi_1,\xi_2,\xi_3,\xi_4) =
   \overline{F^{(1)}(\xi_3,\xi_4,\xi_1,\xi_2)} \\
   F^{(2)}_\epsilon(\xi_1,\xi_2,\xi_3,\xi_4) =
   \overline{F^{(2)}_\epsilon(\xi_4,\xi_3,\xi_2,\xi_1)}
 \end{array}
 $$
and
 $$
   F_\epsilon^{(5)} (\xi_1, \xi_2) =
   \overline{F_\epsilon^{(5)} (\xi_2, \xi_1)},\
   \ \epsilon = +, -
 $$
%%%%%%%%%% DEFINITION OF THE TOTAL HAMILTONIAN  %%%%%%
\begin{defi}\label{def:hamiltonian}
The Hamiltonian for relativistic electrons and po\-si\-trons in a
Cou\-lomb potential interacting with photons in Coulomb gauge that
we consider is given by
 $$
  H(g_1,\, g_2) := H_0 + g_1\, H_I^{(1)}+g_2 H_I^{(2)}\otimes\1 ,
 $$
where $g_i$, $i=1,\,2$, are real coupling constants.
\end{defi}

It is easy to show that $H(g)$ is a symmetric operator in
${\mathcal F}$ as soon as the kernels $F^{(i)}$'s and $G^{(i)}$'s
are square integrable. With stronger conditions on the $F^{(i)}$'s
and $G^{(i)}$'s, we recover a self-adjoint operator with a ground
state, as stated in Theorem~\ref{thm1}~and~\ref{thm2} below.

Let, for $\beta = 0,1,2$,
\begin{eqnarray}\nonumber
&& C_\beta = \sum_{\mu =1,2} \left( \sum_{\gamma,\gamma',n,\ell} \int_{\er^3}
|G_{d,\gamma,\gamma',n,\ell}^\mu(k)|^2 \omega(k)^{-\beta} \d^3k\right)^{1/2}\\
\noalign{\vskip 8pt} &&+ \sum_{\varepsilon = +,-} \sum_{\mu =1,2} \left(
\sum_{\gamma,\gamma',n} \int_{\er^3 \times \er^+}
|G_{d,\varepsilon,\gamma,\gamma',n}^\mu(p;k)|^2 \omega(k)^{-\beta} \d p\,\d^3k
\right)^{1/2} \nonumber\\ \noalign{\vskip 8pt} &&+ \sum_{r} \sum_{\mu =1,2}
\left( \sum_{\gamma,\gamma'} \int_{\er^3 \times \er^3 \times \er^+}
|G_{r,\gamma,\gamma'}^\mu(p,p';k)|^2 \omega(k)^{-\beta} \d p\,\d p'\,\d^3k
\right)^{1/2} \nonumber
\end{eqnarray}
where $r = \{+,+\}, \{+,-\}, \{-,-\}$. We also set
 $$
   F_\epsilon^{(2),a} (\xi_1, \xi_2, \xi_3, \xi_4) : =
   F_\epsilon^{(2)} (\xi_1, \xi_2, \xi_3, \xi_4) -
   F_\epsilon^{(2)} (\xi_1, \xi_2, \xi_4, \xi_3)
 $$

%%%%%%%%%%%%%%%%%%%%%  THEOREM ON SELF-ADJOINTNESS %%%%%%%%%%%%%

We now state our main results.
\begin{theo}\label{thm1} We assume that every $F^{(j)}\in L^2$
($j=1,2,3,4,5,6$). Furthermore, we suppose that $C_0 < \infty$,
$C_1<\infty$ and
 \begin{equation}\label{hypothesis1}
  \displaystyle {\frac{|g_1|}{\sqrt{E_0}}} C_1 +{\frac{|g_2|}{E_0}}
  \left( \frac{1}{\sqrt 2} \|F^{(1)}\| + \|F_+^{(2),a}\| +
  \|F_-^{(2),a}\| \right) <1
 \end{equation}
Then, $H(g_1, g_2)$ is self-adjoint on the domain
$\mathfrak{D}(H_0)$.
\end{theo}

%%%%%%%%%%%%% EXISTENCE OF A GROUND STATE %%%%%%%%%%%%%%%%%%%%%%
\begin{theo}\label{thm2}
We assume that every $F^{(j)}\in L^2$ ($j=1,2,3,4,5,6$), and that
$C_0<\infty$, $C_1 <\infty$ and $C_2 <\infty$. Furthermore, we
suppose that \eqref{hypothesis1} holds true. Then there exists
$g_0>0$ such that for $|g_1|+|g_2| \leq g_0$, the self-adjoint
operator $H(g_1, g_2)$ has a ground state.
\end{theo}

%%%%%%%%%%%%%%%%%%%%%%%%%%%%%%%%%%%%%%%%%%%%%%%%%%%%%%%%%%%%%%%
%%%%%%%%%%  PROOF OF MAIN THEOREMS  %%%%%%%%%%%%%%%%%%%%%%%%%%%
%%%%%%%%%%%%%%%%%%%%%%%%%%%%%%%%%%%%%%%%%%%%%%%%%%%%%%%%%%%%%%%
\section{Proofs of Theorems~\ref{thm1} and \ref{thm2}}\label{S3}

Let
\begin{eqnarray*}
 {\mathcal D}  &=& \{\gamma = (j,m_j,k_j) ; j = \frac{1}{2}, \frac{3}{2},
 \ldots, m_j = -j , \ldots, j, k_j = \pm 1, \pm 2,\ldots,\}\\ \noalign{\vskip
 8pt} {\mathcal D}_d &=& \{(\gamma,n) ; \gamma \in {\mathcal D}
 \ \mbox{and}\ n=0, 1, 2, \ldots \}
\end{eqnarray*}
We set
\begin{eqnarray*}
 \Sigma_+ &=& \Sigma_- = \er^+ \times {\mathcal D}\\
 \noalign{\vskip 8pt} \Sigma_d &=&
 \er^+ \times {\mathcal D}_d
\end{eqnarray*}
>From now on, we will write $\xi = (p,\gamma)$ where $(p,\gamma)$
belongs to $\Sigma_+$ and $\Sigma_-$ respectively and $\xi_d =
(p,\gamma,n)$ where $(p,\gamma,n)$ belongs to $\Sigma_d$. Then
 $$
  L^2(\Sigma_\pm) = \left\{f : \Sigma_\pm \rightarrow \C;
  \ \int_{\Sigma_\pm} |f(\xi)|^2 \d\xi := \Sigma_{\gamma \in {\mathcal D}}
  \int_{\er^+} |f(p,\gamma)|^2 \d p < \infty \right\} .
 $$
Thus the map
 $$
 {(f_\gamma(\cdot))}_{\gamma \in {\mathcal D}} \in \mathop
 \oplus_{\gamma\in {\mathcal D}} L^2(\er^+) \longrightarrow f(\xi) = f(p,\gamma)
 \in L^2(\Sigma_\pm) ,
 $$
is unitary and we will identify the space $\tilde {\mathfrak H}_{c_+}$ (resp.
$\tilde {\mathfrak H}_{c_-}$) introduced in Section~\ref{S2} with
$L^2(\Sigma_+)$(resp. $L^2(\Sigma_-)$).

Similarly we identify the space $\tilde {\mathfrak H}_d$ with a closed
subspace, denoted by $F$, of $L^2(\Sigma_d)$. According to Section~\ref{S2},
$F$ is the closure of the following subspace $$ \left\{ \sum_{(\gamma,n)\in
{\mathcal D}_d} c_{\gamma,n}\, \1_{[n,n+1]}(p)\ ;\ \sum_{(\gamma,n)\in
{\mathcal D}_d} |c_{\gamma,n}|^2 < \infty \right\} $$ Hence the Fermi-Fock
space ${\mathfrak F}_D$ will be identified with $$ {\mathfrak F}_a(F) \otimes
{\mathfrak F}_a(L^2(\Sigma_+)) \otimes {\mathfrak F}_a(L^2(\Sigma_-)) $$ Let us
now define the new annihilation and creation operators. Recall that (See
\cite{ref8})
\begin{eqnarray*}
\left\Vert \int_{\er^+} \d p \ \overline{h_\pm(p,\gamma)}\,
b_{\gamma\pm}(p)\right\Vert &=& \left\Vert \int \d p \, h_\pm(p,\gamma)\,
b_{\gamma,\pm}^*(p)\right\Vert\\ \noalign{\vskip 8pt} &=& \|h_\pm
(\cdot,\gamma)\|_{L^2(\er^+)}
\end{eqnarray*}
Thus the series $\displaystyle \sum_{\gamma \in {\mathcal D}} \int_{\er^+} \d p
\, \overline{h_\pm(p,\gamma)} b_{\gamma,\pm}(p)$ and $\displaystyle
\sum_{\gamma \in {\mathcal D}} \int_{\er^+} \d p \, h_\pm(p,\gamma)
b_{\gamma,\pm}^*(p)$ are normally convergent in $\ell^2({\mathcal D},\
B({\mathfrak F}_a(\tilde {\mathfrak H}_{c\pm}))$ respectively, for every
$h_{\pm}$ in $L^2(\Sigma_{\pm}).$

Here $B({\mathfrak F}_a(\tilde {\mathfrak H}_{c\pm}))$ is the set of bounded
operators in ${\mathfrak F}_a(\tilde {\mathfrak H}_{c\pm})$ respectively.

We then set, for every $h_{\pm}$ in $L^2(\Sigma_{\pm})$ respectively,
\begin{eqnarray*}
b_\pm(h_\pm)   &:=& \int \d\xi\, \overline{h_\pm(\xi)}\, b_\pm(\xi) =
\sum_{\gamma \in {\mathcal D}} \int_{\er^+} \d p\, \overline{h_\pm(p,\gamma)}\,
b_{\gamma,\pm}(p)\\ \noalign{\vskip 8pt} b_\pm^*(h_\pm) &:=& \int \d\xi\,
h_\pm(\xi)\, b_\pm^*(\xi) = \sum_{\gamma \in {\mathcal D}} \int_{\er^+} \d p\,
h_\pm(p,\gamma)\, b_{\gamma,\pm}^*(p)
\end{eqnarray*}
where $b_{\gamma,\pm}(p)$ and $b_{\gamma,\pm}^*(p)$ are defined in
Section~\ref{S2}.

Similarly the series $\displaystyle \sum_{(\gamma,n)\in{\mathcal
D}_d}\!\! \overline{c_{\gamma,n}}\, b_{\gamma,n}$ and
$\displaystyle \sum_{(\gamma,n)\in{\mathcal D}_d}\!\!
c_{\gamma,n}\, b_{\gamma,n}^*$, where $\displaystyle
\sum_{(\gamma,n)\in{\mathcal D}_d}\!\! |c_{\gamma,n}|^2\! <\!
\infty$, are normally convergent in $\ell^2({\mathcal D}_d,
B({\mathfrak F}_a(\tilde {\mathfrak H}_d)))$.

Thus, for any $h_d(\xi_d) = \displaystyle \left( c_{\gamma,n}\,
\1_{[n,n+1]}(p)\right)_{(\gamma,n)\in{\mathcal D}_d} \subset
L^2(\Sigma_d)$, we set
\begin{eqnarray*}
b_d(h_d)   &=& \sum_{(\gamma,n)\in {\mathcal D}_d}
\overline{c_{\gamma,n}}\, b_{\gamma,n} ,\\ \noalign{\vskip 8pt}
b_d^*(h_d)&=& \sum_{(\gamma,n)\in {\mathcal D}_d} c_{\gamma,n}\,
b_{\gamma,n}^* ,
\end{eqnarray*}
where $b_{\gamma,n}$ and $b_{\gamma,n}^*$ are defined in Section~\ref{S221}.

The new canonical anti-commutation relations are the following ones
\begin{eqnarray}\label{eq:20}
\{b_\pm(h_\pm^1),\ b_\pm^*(h_\pm^2)\} = \langle h_\pm^1, h_\pm^2\rangle_{L^2(\Sigma_\pm)}
\end{eqnarray}
\begin{eqnarray}\label{eq:21}
\{b_\pm^\#(h_\pm^1),\ b_\pm^\#(h_\pm^2)\} = 0
\end{eqnarray}
\begin{eqnarray}\label{eq:22}
 \{b_\pm^\#(h_\pm),\   b_d ^\#(h_d)\} = 0
\end{eqnarray}
\begin{eqnarray}\label{eq:23}
\{b_d(h_d^1),\ b_d^*(h_d^2)\} = (h_d^1, h_d^2)_ {L^2(\Sigma_d)}
\end{eqnarray}
Here $b^\#$ is $b$ or $b^*$.

We now have the following lemmas whose proofs are well-known.

\begin{lem}\label{lem2.1}
Let $f$ be in $L^2(\er^3,\C^2;\omega(k)^{-\beta} \d^3k)$ for $\beta = 0,1$. We
have
\begin{eqnarray*}
\|a(f) \Psi\|_{{\mathfrak F}_{ph}}^2&\leq & \left(
\sum_{\mu=1,2}\int \frac{|f(k,\mu)|^2}{\omega(k)}\, \d^3k
\right) \|H_{ph}^{1/2} \Psi\|_{{\mathfrak F}_{ph}}^2\\
\noalign{\vskip 8pt} \|a^*(f) \Psi\|_{{\mathfrak F}_{ph}}^2 &\leq
& \left(\sum_{\mu=1,2} \int \frac{|f(k,\mu)|^2}{\omega(k)}\, \d^3k
\right) \|H_{ph}^{1/2} \Psi\|_{{\mathfrak F}_{ph}}^2 \\ &+& \left(
\sum_{\mu=1,2}\int {|f(k,\mu)|^2}\, \d^3k \right)\,
\|\Psi\|_{{\mathfrak F}_{ph}}^2
\end{eqnarray*}
for every $\Psi \in \D(H_{ph})$ (See \cite{ref4}).
\end{lem}

\begin{lem}\label{lem2.2}
For $h_\pm \in L^2(\Sigma_\pm)$, $b_\pm(h_\pm)$ and
$b_\pm^*(h_\pm)$ are bounded operators and we have $$
\|b_\pm(h)\Psi\|^2 + \|b_\pm^*(h) \Psi\|^2 = \|\Psi\|^2
\|h\|_{L^2(\Sigma_\pm)}^2. $$ In particular, $$ \|b_\pm(h)\| =
\|b_\pm^*(h)\| = \|h\|_{L^2(\Sigma_\pm)} $$ for every $\Psi \in
{\mathfrak F}_a(L^2(\Sigma_\pm))$ respectively (See \cite{ref8}).
\end{lem}
\medskip

\begin{lem}\label{lem2.3}
For every $h_d \in F$, $b_d^\#(h_d)$ is a bounded operator and we have $$
\|b_d(h_d)\Psi\|^2 + \|b_d^*(h_d)\Psi\|^2 = \|h_d\|_{L^2(\Sigma_d)}^2
\|\Psi\|^2 $$ for all $\Psi \in {\mathfrak F}_a(L^2(\Sigma_d))$.
\end{lem}

Let $\{f_i, i=1,2,\ldots\}$ (resp. $\{g_j, j =1,2,\ldots\}$,
$\{h_k, k=1,2,\ldots\}$) be an orthonormal basis of $L^2(F)$
(resp. $L^2(\Sigma_+),\ L^2(\Sigma_-)$). We suppose that the
$g_j$'s and the $h_k$'s are smooth functions in the Schwartz space
with respect to $p$.

We will now consider vectors in ${\mathfrak F}_D$ of the following
form:
\begin{equation}\label{eq:24}
\Phi^{(\ell,m,n)} := b_d^*(f_{i_1}) \ldots b_d^*(f_{i_\ell})
b_+^*(g_{j_1}) \ldots b_+^*(g_{j_m}) b_-^*(h_{k_1}) \ldots
b_-^*(h_{k_n}) \, \Omega ,
\end{equation}
where $l,\ m,\  n$ are positive integers and $\Omega = \Omega_d
\otimes \Omega_{c+} \otimes \Omega_{c-}$. The indexes will be
assumed ordered such that $i_1 < \ldots < i_l$, $j_1 < \ldots <
j_m$ and $k_1 < \ldots < k_n$. It is known that the set
$\{\Phi^{(\ell,m,n)} ; \ell ,m,n= 0, 1, 2, \ldots\}$ is an
orthonormal basis of ${\mathfrak F}_D$ (See \cite{ref8}). The set
 $$
  \mathfrak{F}_{0,D}\! =\! \{ \Psi\in\mathfrak{F}_D;\ \Psi \mbox{ is
  a finite linear combination of basis vectors of the form
  \eqref{eq:24}} \}
 $$
is dense in ${\mathfrak F}_D$.

In the following propositions we investigate several operators in
${\mathfrak F}_D$ built from product of creation operators or
annihilation operators only. In the case of electrons we restrict
ourselves to the electrons in the continuous spectrum. The case of
electrons in the discrete spectrum is easier to deal with.
%%%%%%%%%%%%% LEMMA on B_\pm %%%%%%%%%%%%%%%%%%%%%%%%%%%%%%%%

For $G\in L^2(\Sigma_\pm \times \Sigma_\pm)$ the formal operators
 $$
   \int_{\Sigma_\pm\times \Sigma_\pm} \d\xi_1\,
   \d\xi_2 \overline{G(\xi_1,\xi_2)} b_\pm(\xi_1)
   b_\pm(\xi_2)
 $$
are defined as quadratic forms on
$\mathfrak{F}_{0,D}\times\mathfrak{F}_{0,D}$:
 $$
   \int_{\Sigma_\pm\times \Sigma_\pm} \d\xi_1\,
   \d\xi_2 \langle \Psi,\, \overline{G(\xi_1,\xi_2)} b_\pm(\xi_1)
   b_\pm(\xi_2) \Phi \rangle ,
 $$
where $\Psi,\Phi\in\mathfrak{F}_{0,D}$. Mimicking the proof of
\cite[Theorem X.44]{RS}, we get two operators, denoted by $A_\pm$,
associated with these two forms such that $A_\pm$ are
the unique operators in $\mathfrak{F}_D$ so that
$\mathfrak{F}_{0,D} \subset \mathfrak{D}(A_\pm)$ is a core for
$A_\pm$ and
 $$
  A_\pm = \int_{\Sigma_\pm\times\Sigma_\pm} \d\xi_1\, \d\xi_2
  \overline{G(\xi_1,\xi_2)} b_\pm(\xi_1)
   b_\pm(\xi_2)
 $$
as quadratic forms on
$\mathfrak{F}_{0,D}\times\mathfrak{F}_{0,D}$.

>From now on we denote $A_\epsilon$ the operators $A_\pm$ where
$\epsilon = +,-$, associated with the kernels $G_\epsilon(\xi_1,
\xi_2)$. In the same way, we define as quadratic forms on
$\mathfrak{F}_{0,D}\times\mathfrak{F}_{0,D}$ the operators
$A_{\epsilon \epsilon'}$, $A_{\epsilon \epsilon' \epsilon}$ and
$A_{\epsilon \epsilon \epsilon' \epsilon'}$, where $\epsilon,
\epsilon' = +,-$ and $\epsilon \neq \epsilon'$ as follows
 $$
  A_{\epsilon \epsilon'} = \!\int_{\Sigma_\epsilon\times\Sigma_{\epsilon'}}
   \d\xi_1\, \d\xi_2
  \overline{G_{\epsilon \epsilon'}(\xi_1,\xi_2)} b_\epsilon(\xi_1)
   b_{\epsilon'}(\xi_2) ,
 $$
 $$
  A_{\epsilon \epsilon' \epsilon} =\!
  \int_{\Sigma_\epsilon\times\Sigma_{\epsilon'}\times\Sigma_\epsilon}
   \d\xi_1\, \d\xi_2\, \d\xi_3
  \overline{G_{\epsilon \epsilon'\epsilon}(\xi_1,\xi_2,\xi_3)}
  b_\epsilon(\xi_1)  b_{\epsilon'}(\xi_2)b_\epsilon(\xi_3)  ,
 $$
 $$
  A_{\epsilon \epsilon \epsilon' \epsilon'} =\!
  \int_{\Sigma_\epsilon\times\Sigma_\epsilon\times\Sigma_{\epsilon'}
  \times\Sigma_{\epsilon'}}\!\!\!\!\!\!
   \d\xi_1\, \d\xi_2\, \d\xi_3\, \d\xi_4
  \overline{G_{\epsilon \epsilon \epsilon' \epsilon' }(\xi_1,\xi_2,\xi_3,\xi_4)}
  b_\epsilon(\xi_1)  b_{\epsilon}(\xi_2)b_{\epsilon'}(\xi_3)
  b_{\epsilon'}(\xi_4)  ,
 $$
with $G_{\epsilon\epsilon' }$, $G_{\epsilon \epsilon' \epsilon}$
and $G_{\epsilon \epsilon \epsilon' \epsilon' }$ in $L^2$.

Let $\Phi^{(\ell,m,n)}$ be a vector of the form \eqref{eq:24}. In
addition, for simplicity, assume that $\{ i_1,\ldots, i_\ell \} =
\{1,\ldots,\ell \}$, $\{ j_1,\ldots, j_m \}= \{ 1,\ldots, m \}$
and $\{k_1,\ldots ,k_n\} = \{1,\ldots, n\}$, i.e.
 $$
  \Phi^{(\ell,m,n)} = \prod_{i=1}^\ell b_d^*(f_i) \otimes \prod_{j=1}^m b_+^*(g_j)\otimes
  \prod_{k=1}^n b_-^*(h_k) \Omega .
 $$
We claim that
 \begin{equation}\label{n50}
  A_+ \Phi^{(\ell,m,n)} = \!\!\!\!\sum_{1\leq\alpha < \beta \leq m}
  \!\!\!
  (-1)^{\alpha + \beta}\langle G_+^a,\, g_\alpha\otimes
  g_\beta\rangle \prod_{i=1}^\ell b^*_d(f_i)\otimes\!\!\!\!
  \displaystyle\prod_{\buildrel{j\in\{1,\ldots,m\}}\over{j\neq\alpha,j\neq\beta}}
  \!\!\!b_+^*(g_j)
  \otimes\prod_{k=1}^n b_+^*(h_k)
 \end{equation}
where
 $$
 G_+^a(\xi_1,\xi_2) = G_+(\xi_1,\xi_2) - G_+(\xi_2,\xi_1)
 $$
and
 $$
 \langle G_+^a, g_\alpha\otimes g_\beta\rangle = \int
 \overline{G_+^a(\xi_1, \xi_2)} g_\alpha(\xi_1) g_\beta(\xi_2)
 \d\xi_1\, \d\xi_2 .
 $$
Indeed, the canonical anti-commutation relations (CAR)
\eqref{eq:20}-\eqref{eq:23} yield
\begin{eqnarray}
 \lefteqn{\int\d\xi_2 \overline{G_+(\xi_1,\xi_2)} b_+(\xi_2)
 \Phi^{(\ell,m,n)}} & & \nonumber \\
 & = & (-1)^\ell \sum_{\alpha=1}^m (-1)^{\alpha+1} \langle
 G_+(\xi_1,.), g_\alpha\rangle \prod_{i=1}^\ell b_d^*(h_i) \otimes
 \!\!\prod_{\buildrel{j\in\{1,\ldots,m\}}\over{j\neq\alpha}}\!\!b_+^*(g_j)
 \otimes\prod_{k=1}^n b_-^*(h_k) \Omega\label{n51}
\end{eqnarray}

We have the identity $\langle G_+(\xi_1, .), g_\alpha\rangle =
\sum_{\beta=1}^\infty \langle G_+, g_\beta\otimes g_\alpha\rangle
\overline{g_\beta(\xi_1)}$ where $\langle G_+, g_\beta\otimes
g_\alpha\rangle = \int_{\Sigma_+\times\Sigma_+}
\overline{G_+(\xi_1, \xi_2)} g_\beta(\xi_1) g_\alpha(\xi_2)
\d\xi_1\, \d\xi_2$.

Applying $b_+(\xi_1)$ to the right hand side
of \eqref{n51}, using the CAR as above and integrating with
respect to $\xi_1$ we get
\begin{equation}\label{n52}
 \begin{split}
  A_+\Phi^{(\ell,m,n)}
  = \sum_{\alpha=1}^{m-1} (-1)^{\alpha +1}\!\!\!\!\!
  \sum_{1\leq \alpha < \beta \leq m}\!\!\! \langle G_+, g_\alpha\!\otimes
  \!g_\beta \rangle\\
  \ \ (-1)^{\beta +1} \prod_{i=1}^\ell b_d^*(f_i) \otimes
  \prod_{\buildrel{j\in\{1,\ldots,m\}}\over{j\neq\alpha, j\neq\beta}}
  b_+^*(g_j)\otimes \prod_{k=1}^n b_+^*(h_k)\Omega\\
  + \sum_{\alpha=1}^{m-1} (-1)^{\alpha +1} \sum_{1\leq \alpha
  <\beta \leq m} \langle G_+, g_\beta\otimes g_\alpha\rangle\\
  (-1)^\beta \prod_{i=1}^\ell b_d^*(f_i)\otimes
  \prod_{\buildrel{j\in\{1,\ldots,m\}}\over{j\neq\alpha,
  j\neq\beta}}b_+^*(g_j)\otimes \prod_{k=1}^n b_-^*(h_k)\Omega ,
 \end{split}
\end{equation}
from which we deduce \eqref{n50}. Similarly, we have
\begin{equation}\label{n53}
\begin{split}
 A_-\Phi^{(\ell,m,n)} & = \sum_{1\leq \gamma <\delta \leq n}
 (-1)^{\gamma + \delta} \langle G_-^a, h_\gamma\otimes
 h_\delta\rangle \\
 & \prod_{i=1}^\ell b_d^*(f_i)\otimes \prod_{j=1}^m
 b_+^*(g_j) \otimes \prod_{\buildrel{k\in\{1,\ldots,n\}} \over{k\neq\gamma,
  k\neq\delta}} b_+^*(h_k)\Omega ,
\end{split}
\end{equation}
where $$G_-^a(\xi_1, \xi_2) = G_- (\xi_1, \xi_2) - G_-(\xi_2,
\xi_1)$$ and $\langle G_-^a, h_\gamma\otimes h_\delta\rangle =
\int_{\Sigma_-\times\Sigma_-} \overline{G_-^a(\xi_1, \xi_2)}
h_\gamma(\xi_1) h_\delta(\xi_2) \d\xi_1\, \d\xi_2$.

In a complete similar way, we also get
\begin{equation}\label{n54}
\begin{split}
A_{+-} \Phi^{(\ell,m,n)} = & (-1)^m \sum_{\alpha =1}^m
\sum_{\beta=1}^n (-1)^{\alpha + \beta} \langle G_{+-},
g_\alpha\otimes h_\beta\rangle \\ & \prod _{i=1}^\ell b_d^*(f_i)
\otimes \prod_{\buildrel{j\in\{1,\ldots,m\}}\over{j\neq\alpha}}
b_+^*(g_j)\otimes\prod_{\buildrel{k\in\{1,\ldots,n\}}\over{k\neq\beta}}
b_-^*(h_k) \Omega
\end{split}
\end{equation}
\begin{equation}\nonumber
\begin{split}
A_{-+} \Phi^{(\ell,m,n)} = & (-1)^m \sum_{\alpha =1}^m
\sum_{\beta=1}^n (-1)^{\alpha + \beta} \langle G_{-+},
h_\beta\otimes g_\alpha\rangle \\ & \prod _{i=1}^\ell b_d^*(f_i)
\otimes \prod_{\buildrel{j\in\{1,\ldots,m\}}\over{j\neq\alpha}}
b_+^*(g_j)\otimes\prod_{\buildrel{k\in\{1,\ldots,n\}}\over{k\neq\beta}}
b_-^*(h_k) \Omega
\end{split}
\end{equation}
where
 $$
   \langle G_{+-}, g_\alpha\otimes h_\beta\rangle =
   \int_{\Sigma_+\times \Sigma_-} \overline{G_{+-}(\xi_1,\xi_2)}
   g_\alpha(\xi_1) h_\beta(\xi_2) \d\xi_1\, \d\xi_2
 $$
and
 $$
   \langle G_{-+}, h_\beta\otimes g_\alpha\rangle =
   \int_{\Sigma_-\times \Sigma_+} \overline{G_{-+}(\xi_1,\xi_2)}
   h_\beta(\xi_1) g_\alpha(\xi_2) \d\xi_1\, \d\xi_2 .
 $$
Using \eqref{n54} we get, as for \eqref{n52}
\begin{equation}\label{n56}
 \begin{split}
  A_{-+-} \Phi^{(\ell,m,n)} = &\sum_{\alpha =1}^m \sum_{\beta =1}^n
  (-1)^{\alpha +\beta}\sum_{1\leq \gamma < \beta \leq n}
  (-1)^{\gamma+1} \langle G_{-+-}^a, h_\gamma\otimes
  g_\alpha\otimes h_\beta\rangle \\
  & \prod_{i=1}^\ell b_d^*(f_i) \otimes
  \prod_{\buildrel{j\in\{1,\ldots,m\}}\over{j\neq\alpha}}
  b_+^*(g_j) \otimes \prod_{\buildrel{k\in\{1,\ldots,n\}}\over{k\neq\beta,
  k\neq\gamma}}b_-^*(h_k)\Omega
 \end{split}
\end{equation}
and
\begin{equation}\nonumber
 \begin{split}
  A_{+-+} \Phi^{(\ell,m,n)} = &\sum_{\alpha =1}^m \sum_{\beta =1}^n
  (-1)^{\alpha +\beta}\sum_{1\leq \gamma < \beta \leq n}
  (-1)^{\gamma+1} \langle G_{+-+}^a, g_\gamma\otimes
  h_\alpha\otimes g_\beta\rangle \\
  & \prod_{i=1}^\ell b_d^*(f_i) \otimes
  \prod_{\buildrel{j\in\{1,\ldots,m\}}\over{j\neq\gamma, j\neq\beta}}
  b_+^*(g_j) \otimes \prod_{\buildrel{k\in\{1,\ldots,n\}}
  \over{k\neq\alpha}}b_-^*(h_k)\Omega
 \end{split}
\end{equation}
where
 $$
  G_{\epsilon \epsilon' \epsilon}^a (\xi_1,\xi_2,\xi_3) =
  G_{\epsilon \epsilon' \epsilon} (\xi_1,\xi_2,\xi_3) -
  G_{\epsilon \epsilon' \epsilon} (\xi_3,\xi_2,\xi_1)
 $$
and
 $$
   \langle G^a_{-+-}, h_\gamma\otimes g_\alpha \otimes
   h_\beta\rangle =
   \int_{\Sigma_-\times \Sigma_+ \times \Sigma_-}\!\!\!\!
   \overline{G^a_{-+-}(\xi_1, \xi_2, \xi_3)}h_\gamma(\xi_1)
   g_\alpha(\xi_2) h_\beta(\xi_3) \d\xi_1\, \d\xi_2\, \d\xi_3
 $$
 $$
  \langle G^a_{+-+}, g_\gamma\otimes h_\alpha \otimes
   g_\beta\rangle =
   \int_{\Sigma_+\times \Sigma_- \times \Sigma_+}\!\!\!\!
   \overline{G^a_{+-+}(\xi_1, \xi_2, \xi_3)}g_\gamma(\xi_1)
   h_\alpha(\xi_2) g_\beta(\xi_3) \d\xi_1\, \d\xi_2\, \d\xi_3
 $$
By \eqref{n50} and \eqref{n53}, we finally obtain
\begin{equation}\label{n58}
\begin{split}
  A_{++--} \Phi^{(\ell,m,n)} = & \sum_{1\leq \alpha < \beta\leq m}
 (-1)^{\alpha + \beta} \sum_{1 \leq \gamma < \delta \leq n}
 (-1)^{\gamma + \delta} \langle G_{++--}^{aa},
 g_\alpha\otimes g_\beta\otimes h_\gamma\otimes h_\delta\rangle\\
 & \prod_{i=1}^\ell b_d^*(f_i) \otimes \prod_{\buildrel{j\in\{1,\ldots,m\}}
  \over{j\neq\alpha, j\neq\beta}} b_+^*(g_j) \otimes
  \prod_{\buildrel{k\in\{1,\ldots,n\}}
  \over{k\neq\gamma, k\neq\delta}}b_-^*(h_k) \Omega
\end{split}
\end{equation}
and
\begin{equation}\nonumber
\begin{split}
  A_{--++} \Phi^{(\ell,m,n)} = & \sum_{1\leq \gamma < \delta\leq n}
 (-1)^{\gamma + \delta} \sum_{1 \leq \alpha < \beta \leq m}
 (-1)^{\alpha + \beta} \langle G_{--++}^{aa},
 h_\gamma\otimes h_\delta \otimes g_\alpha\otimes g_\beta\rangle\\
 & \prod_{i=1}^\ell b_d^*(f_i) \otimes \prod_{\buildrel{j\in\{1,\ldots,m\}}
  \over{j\neq\alpha, j\neq\beta}} b_+^*(g_j) \otimes
  \prod_{\buildrel{k\in\{1,\ldots, n\}}
  \over{k\neq\gamma, k\neq\delta}}b_-^*(h_k) \Omega
\end{split}
\end{equation}
where
\begin{eqnarray*}
  G_{\epsilon\epsilon \epsilon'
  \epsilon'}^{aa}(\xi_1,\xi_2,\xi_3,\xi_4)
   &= & G_{\epsilon\epsilon \epsilon'
  \epsilon'}(\xi_1,\xi_2,\xi_3,\xi_4)
  -  G_{\epsilon\epsilon \epsilon'
  \epsilon'}(\xi_1,\xi_2,\xi_4,\xi_3)\\
  & & - G_{\epsilon\epsilon \epsilon'
  \epsilon'}(\xi_2,\xi_1,\xi_3,\xi_4)
  +  G_{\epsilon\epsilon \epsilon'
  \epsilon'}(\xi_2,\xi_1,\xi_4,\xi_3) .
\end{eqnarray*}

We have the following proposition
\begin{prop}\label{nprop3.4}
We have, for $\epsilon\neq\epsilon'$
\begin{itemize}
 \item[{(i)}] $\displaystyle\|A_\epsilon\| = \|A_\epsilon^*\| \leq
 \|G_\epsilon^a\|_{L^2(\Sigma_\epsilon\times \Sigma_\epsilon)} $

 \item[{(ii)}] $\displaystyle\|A_{\epsilon\epsilon'}\| =
 \|A^*_{\epsilon\epsilon'}\| \leq \|G_{\epsilon
 \epsilon'}\|_{L^2(\Sigma_\epsilon\times\Sigma_{\epsilon'})}$

 \item[{(iii)}] $\displaystyle\|A_{\epsilon\epsilon'\epsilon}\| =
 \|A^*_{\epsilon\epsilon'\epsilon}\| \leq \|G^a_{\epsilon
 \epsilon'\epsilon}\|_{L^2(\Sigma_\epsilon\times\Sigma_{\epsilon'}
 \times\Sigma_\epsilon)}$

 \item[{(iv)}] $\displaystyle\|A_{\epsilon\epsilon\epsilon'\epsilon'}\| =
 \|A^*_{\epsilon\epsilon\epsilon'\epsilon'}\| \leq \|G^{aa}_{\epsilon
 \epsilon\epsilon'\epsilon'}\|_{L^2(\Sigma_\epsilon\times\Sigma_{\epsilon}
 \times\Sigma_{\epsilon'}\times\Sigma_{\epsilon'})}$
\end{itemize}
\end{prop}
\begin{rema}
When we consider electrons in the discrete spectrum, the estimates
of Proposition~\ref{nprop3.4} are still true. We just have to
substitute $\Sigma_d$ for $\Sigma_+$.
\end{rema}
\begin{proof}
As mentioned above, for the proof in the case of electrons, we
only consider electrons in the continuous spectrum. The proof for
electrons in the discrete spectrum is similar and simpler. For the
proof of (i) we restrict ourselves to $A_+$. The case of $A_-$ is
similar. Since $\displaystyle\left(\prod_{i=1}^\ell
b_d^*(f_i)\otimes \prod_{\buildrel{j\in\{1,\ldots,m\}}
\over{j\neq\alpha, j\neq\beta}}b_+^*(g_j)\otimes\prod_{k=1}^n
b_+^*(h_k) \Omega \right)_{1\leq\alpha <\beta\leq m}$ is an
orthonormal family, it follows from \eqref{n50} that
\begin{equation}\nonumber
%\begin{split}
 \|A_+ \Phi^{(\ell,m,n)}\|^2  =\!\! \sum_{1 \leq \alpha < \beta \leq m}
 |\la G_+, g_\alpha \otimes g_\beta\rangle
 |^2_{L^2(\Sigma_+\times\Sigma_+)} \\
  \leq \|G_+^a\|^2_{L^2(\Sigma_+\times\Sigma_+)}
 \|\Phi^{(\ell,m,n)}\|^2 .
%\end{split}
\end{equation}
In order to prove (i), it is enough to show that
\begin{equation}\label{n61}
 \|A_+ \Psi\|^2 \leq \|G^a_+\|^2 \|\Psi\|^2
\end{equation}
for every $\Psi\in\mathfrak{F}_{0,D}$.

The most significant finite linear combination of basis vectors is
the following one
\begin{equation}\label{n62}
\Psi =
\sum_{\mu=1}^N\sum_{\nu=1}^P\lambda_{\mu\nu}\prod_{i=1}^\ell
b_d^*(f_i)\otimes \prod_{j=1}^{m-1} b_+^*(g_j)
b_+^*(g_{m_\mu})\otimes \prod_{k=1}^{n-1}b_-^*(h_k)
b_-^*(h_{n_\nu}) \Omega
\end{equation}
Here $N$ and $P$ are positive integers and we have $m-1
<m_1<m_2<\ldots<m_N$ and $ n-1<n_1<n_2<\ldots <n_P$. From  now on
we restrict ourselves to finite linear combination of the form
\eqref{n62}. The proof of inequality \eqref{n61} is easier for any
other finite linear combination of basis vectors. Note that
$\|\Psi\|^2 = \sum_{\mu=1}^N \sum_{\nu=1}^P |\lambda_{\mu\nu}|^2$.
By \eqref{n50} we get
\begin{equation}\label{n63}
\begin{split}
 A_+\Psi = & \sum_{\mu=1}^N \sum_{\nu=1}^P \sum_{1\leq
 \alpha<\beta\leq m-1} (-1)^{\alpha + \beta}
 \lambda_{\mu\nu} \langle G_+^a, g_\alpha \otimes g_\beta\rangle\\
 & \prod_{i=1}^\ell b_d^*(f_i) \otimes
  \prod_{\buildrel{j\in\{1,\ldots,m-1\}}
  \over{j\neq\alpha, j\neq\beta}}b_+^*(g_j)
  b_+^*(g_{m_\mu})\otimes \prod_{k=1}^{n-1}
  b_-^*(h_k) b_-^*(h_{n_\nu})\Omega\\
  & + \sum_{\nu=1}^P\sum_{1\leq\alpha\leq m-1}
  (-1)^{\alpha+m}\left(\sum_{\mu=1}^N \lambda_{\mu\nu}\langle G_+^a,
  g_\alpha\otimes g_{m_\mu}\rangle \right)\\
  & \prod_{i=1}^\ell b_d^*(f_i) \otimes  \prod_{\buildrel{j\in\{1,\ldots,m-1\}}
  \over{j\neq\alpha}}b_+^*(g_j)\otimes \prod_{k=1}^{n-1}
  b_-^*(h_k) b_-^*(h_{n_\nu})\Omega  .
\end{split}
\end{equation}
The right hand side of \eqref{n63} is a linear combination of
vectors of an orthogonal family. Thus
\begin{equation}\nonumber
\begin{split}
\|A_+\Psi\|^2 = & \sum_{1\leq\alpha<\beta\leq m-1}
(\sum_{\nu=1}^P\sum_{\mu=1}^N |\lambda_{\mu\nu}|^2) |\langle
G_+^a, g_\alpha \otimes g_\beta \rangle|^2 \\ & +
\sum_{1\leq\alpha\leq m-1} \sum_{\nu=1}^P |\sum_{\mu=1}^N
\lambda_{\mu\nu} \langle G_+^a, g_\alpha\otimes g_{m_\mu}\rangle
|^2
\end{split}
\end{equation}
By the Cauchy-Schwarz inequality we get
\begin{equation}\nonumber
\begin{split}
\|A_+\Psi\|^2 \leq &(\sum_{\mu=1}^N\sum_{\nu=1}^P
|\lambda_{\mu\nu}|^2) \Big[ \sum_{1\leq\alpha<\beta\leq m-1}
\left| \langle G_+^a, g_\alpha\otimes g_\beta\rangle\right|^2 \\
 & + \sum_{1\leq\alpha\leq m-1}\sum_{\mu=1}^N |\langle G_+^a,
 g_\alpha\otimes g_{m_\mu}\rangle |^2 \Big] \leq \|G_+^a\|^2
 \|\Psi\|^2 .
\end{split}
\end{equation}
This concludes the proof of (i). Since
 $$
  \displaystyle\left(\prod_{i=1}^\ell b_d^*(f_i)\otimes
  \prod_{\buildrel{j\in\{1,\ldots,m\}}
  \over{j\neq\alpha}}b_+^*(g_j)\otimes
  \prod_{\buildrel{k\in\{1,\ldots,n\}} \over{k\neq\beta}} b_+^*(h_k)
  \Omega \right)_{\buildrel{1\leq\alpha\leq m}\over{1\leq\beta\leq
  n}}
 $$
is an orthonormal family we have
\begin{equation}\nonumber
 \|A_{+-} \Phi^{(\ell,m,n)}\|^2 = \sum_{\alpha=1}^m
 \sum_{\beta=1}^n |\langle G_{+-},g_\alpha\otimes h_\beta\rangle|^2
 \leq\|G_{+-}\|^2 \|\Phi^{(\ell,m,n)}\|^2
\end{equation}
Let $\Psi$ be of the form \eqref{n62}. By \eqref{n54} we get
\begin{equation}\label{n67}
\begin{split}
 {A_{+-}\Psi}
  & =  (-1)^m \sum_{\mu=1}^N \sum_{\nu=1}^P
 \sum_{\alpha=1}^{m-1}\sum_{\beta=1}^{n-1} (-1)^{\alpha + \beta}
 \lambda_{\mu\nu}\langle G_{+-}, g_\alpha\otimes h_\beta\rangle\\
 & \prod_{i=1}^\ell b_d^*(f_i)\otimes\!\!\!\!\!
 \prod_{\buildrel{j\in\{1,\ldots,m-1\}}
 \over{j\neq\alpha}}\!\!\!b_+^*(g_j) b_+^*(g_{m_\mu})\otimes
 \!\!\!\!\!\prod_{\buildrel{k\in\{1,\ldots,n-1\}}
 \over{k\neq\beta}} \!\!\!b_-^*(h_k) b_-^*(h_{n_\nu})\Omega\\
 & + (-1)^m \sum_{\mu=1}^N\sum_{\beta=1}^{n-1} (-1)^\beta \!
 \sum_{\nu=1}^P\lambda_{\mu\nu} \langle G_{+-}, g_{m_\mu}\otimes
 h_\beta\rangle\\
 & \prod_{i=1}^\ell b_d^*(f_i)\otimes \prod_{j=1}^{m-1} b_+^*(g_j)
 \otimes \prod_{\buildrel{k\in\{1,\ldots,n-1\}}
 \over{k\neq\beta}}b_-^*(h_k) b_-^*(h_{n_\nu})\Omega \\
 & + (-1)^{m+n} \sum_{\nu=1}^P\sum_{\alpha=1}^{m-1} (-1)^\alpha
 \sum_{\mu=1}^N\lambda_{\mu\nu}\langle G_{+-},
 g_\alpha\otimes h_{n_\nu}\rangle \\
 & \prod_{i=1}^\ell
 b_d^*(f_i)\otimes \prod_{\buildrel{j\in\{1,\ldots,m-1\}}
 \over{j\neq\alpha}}b_+^*(g_j)b_+^*(g_{m_\mu})\otimes
 \prod_{k=1}^{n-1}b_-^*(h_k)\Omega \\
 & + (-1)^n \left(\sum_{\mu=1}^N \sum_{\nu=1}^P \lambda_{\mu\nu} \langle
 G_{+-}, g_{m_\mu}\otimes h_{n_\nu}\rangle
 \right)
 \prod_{i=1}^\ell b_d^*(f_i)\otimes\prod_{j=1}^{m-1}b_+^*(g_j)\otimes
 \prod_{k=1}^{n-1}b_-^*(h_k)\Omega
\end{split}
\end{equation}
The right hand side of \eqref{n67} is a linear combination of
vectors of an orthogonal family. Thus we obtain
\begin{eqnarray*}
\lefteqn{\|A_{+-}\Psi\|^2 } & & \\
 & = &\sum_{\mu=1}^N
\sum_{\nu=1}^P\sum_{\alpha=1}^{m-1}\sum_{\beta=1}^{n-1}
|\lambda_{\mu \nu}|^2 |\langle G_{+-}, g_\alpha\otimes h_\beta
\rangle|^2 + \sum_{\beta=1}^{n-1} \sum_{\nu=1}^P
\Big|\sum_{\mu=1}^N \lambda_{\mu\nu} \langle G_{+-},
g_{m_\mu}\otimes h_\beta\rangle\Big|^2 \\  & & +
\sum_{\alpha=1}^{m-1} \sum_{\mu=1}^N \Big| \sum_{\mu=1}^P
\lambda_{\mu\nu} \langle G_{+-}, g_\alpha\otimes h_{n_\nu}\rangle
\Big|^2 + \Big| \sum_{\mu=1}^N \sum_{\nu =1}^P \lambda_{\mu\nu}
\langle G_{+-}, g_{m_\mu}\otimes h_{n_\nu}\rangle \Big|^2 .
\end{eqnarray*}
By the Cauchy-Schwarz
inequality we get
\begin{equation}\nonumber
\begin{split}
 \lefteqn{\|A_{+-}\Psi\|^2} &\\
 & \!\leq\!  \sum_{\mu=1}^N\sum_{\nu=1}^P
 |\lambda_{\mu\nu}|^2\!
 \sum_{\alpha=1}^{m-1}\sum_{\beta=1}^{n-1} |\langle G_{+-},
 g_\alpha\!\otimes\! h_\beta\rangle|^2\!
  +\! \sum_{\beta=1}^{n-1}\sum_{\nu=1}^P \sum_{\mu=1}^N
 |\lambda_{\mu\nu}|^2\! \sum_{\mu=1}^N |\langle
 G_{+-}, g_{m_\mu}\!\otimes\! h_\beta\rangle |^2\\
 & +\!\! \sum_{\alpha=1}^{m-1}\sum_{\mu=1}^N\sum_{\nu=1}^P
 \!|\lambda_{\mu\nu}|^2\!\sum_{\nu=1}^P|\langle G_{+-},
 g_\alpha\!\otimes\! h_{n_\nu}\rangle|^2
  \!+\!\! \sum_{\mu=1}^N\sum_{\nu=1}^P
 \!|\lambda_{\mu\nu}|^2\!\sum_{\mu=1}^N\sum_{\nu=1}^P
 |\langle G_{+-}, g_{m_\mu}\!\otimes\! h_{n_\nu}\rangle |^2\\
 &\leq  \sum_{\mu=1}^N\sum_{\nu=1}^P
 |\lambda_{\mu\nu}|^2 \Bigg[
 \sum_{\alpha=1}^{m-1}\sum_{\beta=1}^{n-1} |\langle G_{+-},
 g_\alpha\otimes h_\beta\rangle|^2 + \sum_{\beta=1}^{n-1}
 \sum_{\mu=1}^N |\langle
 G_{+-}, g_{m_\mu}\otimes h_\beta\rangle |^2\\
 & + \sum_{\alpha=1}^{m-1}\sum_{\nu=1}^P |\langle G_{+-},
 g_\alpha\otimes h_{n_\nu}|^2 + \sum_{\mu=1}^N \sum_{\nu=1}^P
 |\langle G_{+-}, g_{m_\mu}\otimes h_{n_\nu}\rangle |^2
 \Bigg] \leq \|G_{+-}\|^2 \|\Psi\|^2 .
\end{split}
\end{equation}
The proof for $A_{-+}$ is the same. This concludes the proof of
(ii).

Using \eqref{n56} and \eqref{n58} and following the same method as
for proving (i) and (ii), we prove (iii) and (iv). See \cite{bdg2}
for details.

Proposition~\ref{nprop3.4} is thus proved.

\end{proof}

\begin{rema}
The following formal operators
 $$
  \int\d\xi_1\, \d\xi_2\, G_\epsilon(\xi_1, \xi_2)
  b_\epsilon^*(\xi_1) b_\epsilon^*(\xi_2) ,
 $$
 $$
  \int\d\xi_1\, \d\xi_2\, G_{\epsilon\epsilon'}(\xi_1, \xi_2)
  b_\epsilon^*(\xi_1) b_{\epsilon'}^*(\xi_2) ,
 $$
 $$
  \int\d\xi_1\, \d\xi_2\, \d\xi_3\, G_{\epsilon\epsilon'\epsilon}
  (\xi_1, \xi_2, \xi_3)
  b_\epsilon^*(\xi_1) b_{\epsilon'}^*(\xi_2) b_\epsilon^*(\xi_3) ,
 $$
and
 $$
  \int\d\xi_1\, \d\xi_2\, \d\xi_3\, \d\xi_4\,
  G_{\epsilon\epsilon\epsilon'\epsilon'}(\xi_1, \xi_2, \xi_3, \xi_4)
  b_\epsilon^*(\xi_1) b_{\epsilon}^*(\xi_2) b_{\epsilon'}^*(\xi_3)
  b_{\epsilon'}^*(\xi_4)  ,
 $$
are the ones associated respectively with $A_\epsilon^*$,
$A_{\epsilon, \epsilon'}^*$, $A_{\epsilon, \epsilon',\epsilon}^*$
and $A_{\epsilon\epsilon\epsilon'\epsilon'}^*$, as quadratic forms
on $\mathfrak{F}_{0,D}\times \mathfrak{F}_{0,D}$.
\end{rema}

We now investigate operators in $\mathfrak{F}_D$ built from a
product of creation and annihilation operators. Let us introduce
 \begin{equation}\nonumber
 \begin{split}
  \mathfrak{F}_{D,fin} = \Big\{ \Psi = (\Psi^{(q,r,s)})_{q\geq0,
  r\geq 0, s\geq 0};\ \Psi^{(q,r,s)} \mbox{ is in the Schwartz space and } \\
  \Psi^{(q,r,s)} = 0\ \mbox{for all but
  finitely many }(q,r,s) \Big\} .
 \end{split}
 \end{equation}
$\mathfrak{F}_{D,fin}$ is a core for $N_D$. For $G_\epsilon\in
L^2$,
%the following formal operators
% $$
%  \int_{\Sigma_\epsilon\times\Sigma_\epsilon}\d\xi_1\,\d\xi_2
%  G_\epsilon(\xi_1, \xi_2) b_\epsilon^*(\xi_1) b_\epsilon(\xi_2)
% $$
%are defined as quadratic forms on
%$\mathfrak{F}_{D,fin}\times\mathfrak{F}_{D,fin}$ in the following
%way, for $\Psi, \Phi\in \mathfrak{F}_{D,fin}$
%\begin{equation}\label{n78}
%\int\d\xi_1\,\d\xi_2 \langle b_\epsilon(\xi_1)\Psi,
%G_\epsilon(\xi_1, \xi_2)b_\epsilon(\xi_2)\Phi\rangle
%\end{equation}
by mimicking the proof of \cite[Theorem X.44]{RS}, one can show
that there exists two operators, denoted by $B_\epsilon$
($\epsilon  \in\{ +,-\}$), such that $B_\epsilon$ are the unique
operators in $\mathfrak{F}_D$ such that
$\mathfrak{F}_{D,fin}\subset \mathfrak{D}(B_\epsilon)$ is a core
for $B_\epsilon$ and
\begin{equation}\nonumber
  B_\epsilon = \int_{\Sigma_\epsilon\times\Sigma_\epsilon}\d\xi_1\,\d\xi_2
  G_\epsilon(\xi_1, \xi_2) b_\epsilon^*(\xi_1) b_\epsilon(\xi_2)
\end{equation}
and
\begin{equation}\label{n80}
  B_\epsilon^* = -\int_{\Sigma_\epsilon\times\Sigma_\epsilon}\d\xi_1\,\d\xi_2
  \overline{G_\epsilon(\xi_1, \xi_2)} b_\epsilon^*(\xi_1) b_\epsilon(\xi_2)
\end{equation}
as quadratic forms on
$\mathfrak{F}_{D,fin}\times\mathfrak{F}_{D,fin}$. Similarly to
\eqref{n80}, for $G_{d,\epsilon} =
\left(G_{\gamma,n;\epsilon}\right)_{(\gamma,n)\in\mathcal{D}_d}\in
L^2(\mathcal{D}_d\times \Sigma_\epsilon )$ and $G_{d,d} =
\left(G_{\gamma,n;\gamma',n'}
\right)_{(\gamma,n;\gamma',n')\in\mathcal{D}_d\times\mathcal{D}_d}\in
L^2(\mathcal{D}_d\times\mathcal{D}_d)$, we define the operators
 $$
   B_{d, \epsilon} = \sum_{\gamma, n} \int_{\Sigma_\epsilon}
   \d \xi G_{\gamma, n; \epsilon}(\xi)\, b^*_{\gamma,n} b_\epsilon(\xi)
 $$
and
 $$
   B_{d,d} = \sum_{\gamma, n} \sum_{\gamma', n'} G_{\gamma,n;
   \gamma', n'}b^*_{\gamma,n} b_{\gamma',n'}
 $$

We then have the following proposition whose proof is borrowed
from \cite{GJ}
\begin{prop}\label{prop2.5}
For $G_\epsilon\in L^2(\Sigma_\epsilon\times\Sigma_\epsilon)$,
$G_{d,\epsilon}  \in L^2(\mathcal{D}_d \times \Sigma_\epsilon)$
and $G_{d,d}\in L^2(\mathcal{D}_d\times\mathcal{D}_d)$ we have
\begin{equation}\label{eq:31}
  \|B_\epsilon\, (N_D+1)^{-1/2}\| \leq
  \|G_\epsilon\|_{L^2(\Sigma_\epsilon\times\Sigma_\epsilon)} ,
\end{equation}
\begin{equation}\label{eq:32}
  \|(N_D+1)^{-1/2}\, B_\epsilon\| \leq
  \|G_\epsilon\|_{L^2(\Sigma_\epsilon\times\Sigma_\epsilon)} ,
\end{equation}
\begin{equation}\label{eq:31.bis}
  \|B_{d,\epsilon}\| \leq
  \|G_{d,\epsilon}\|_{L^2(\mathcal{D}_d\times \Sigma_\epsilon)} ,
\end{equation}
and
\begin{equation}\label{eq:31.ter}
  \|B_{d,d}\| \leq
  \|G_{\gamma,n;\gamma', n'}
  \|_{L^2(\mathcal{D}_d\times \mathcal{D}_d)},
\end{equation}

\end{prop}

\begin{proof}
Since $N_D$ is a self-adjoint operator, \eqref{eq:32} follows from
\eqref{n80} and \eqref{eq:31}. We only investigate $B_+$ since the
proof for $B_-$ is the same.

Let $\Psi= (\Psi^{(q,r,s)})$ and $\Phi=(\Phi^{(q',r',s')}$ be two
vectors in $\mathfrak{F}_{D,fin}$. We have
\begin{equation}\nonumber
\begin{split}
\lefteqn{\langle \Phi^{(q',r',s')}, B_+
\Psi^{(q,r,s)}\rangle_{\mathfrak{F}_D}}&
\\
 & \ \ \ =
\delta_{qq'}\delta_{rr'}\delta_{s s'}\int_{\Sigma_+\times\Sigma_+}
\!\!\!G_+(\xi_1,\xi_2) \langle b_+(\xi_1) \Phi^{(q,r,s)},
b_+(\xi_2) \Psi^{(q,r,s)}\rangle_{\mathfrak{F}_a^{(q,r-1,s)}}
\d\xi_1\, \d\xi_2
\end{split}
\end{equation}
Thus $B_+ \Psi^{(q,r,s)}\in\mathfrak{F}_a^{(q,r,s)}$ for every
triple $(q,r,s)$. Therefore, we only need to estimate $\langle
\Phi^{(q,r,s)}, B_+ \Psi^{(q,r,s)}\rangle$ for every triple
$(q,r,s)$. By the Fubini theorem we have
\begin{equation}\nonumber
 |\langle \Phi^{(q,r,s)}, B_+\Psi^{(q,r,s)}\rangle |^2 =
 \left|\int_{\Sigma_+} \left\langle b_+(\xi_1) \Phi^{(q,r,s)},
 \int_{\Sigma_+} G_+(\xi_1, \xi_2)
 b_+(\xi_2) \Psi^{(q,r,s)} \d\xi_2\right\rangle \d\xi_1\right|^2
\end{equation}
By the Cauchy-Schwarz inequality and the fact that $\|b_+(f)\| =
\|f\|$, we get
\begin{equation}\nonumber
 |\langle \Phi^{(q,r,s)}, B_+ \Psi^{(q,r,s)}\rangle|^2 \!\leq\!\!
 \left(\int_{\Sigma_+}\!\!\!\! \|b_+(\xi_1) \Phi^{(q,r,s)}\|\!
 \left(\int_{\Sigma_+}\!\!\!\! |G_+(\xi_1, \xi_2)|^2 \d\xi_2\right)^\frac12
 \!\!\!\d\xi_1\!\!\right)^2\!\!\! \|\Psi^{(q,r,s)}\|^2
\end{equation}
Applying again Cauchy-Schwarz inequality and the definition of
$b_+(\xi)$, we finally get
\begin{equation}
\begin{split}
 |\langle \Phi^{(q,r,s)}, B_+ \Psi^{(q,r,s)}\rangle |^2 &\leq
 r \|G_+\|^2 \| \Phi^{(q,r,s)}\|^2 \|\Psi^{(q,r,s)}\|^2 \\
 & \leq
\|G_+\|^2 \| \Phi^{(q,r,s)}\|^2
\|(N_D+1)^\frac12\Psi^{(q,r,s)}\|^2 .
\end{split}
\end{equation}
Since $B_+\Psi^{(q,r,s)}\in\mathfrak{F}_a^{(q,r,s)}$, we have
\begin{equation}\label{n87}
|\langle\Phi, B_+\Psi^{(q,r,s)}\rangle|^2 \leq \|G_+\|^2 \|
\Phi\|^2 \|(N_D+1)^\frac12\Psi^{(q,r,s)}\|^2 ,
\end{equation}
for every $\Phi\in\mathfrak{F}_{D,fin}$. Now, since
$\mathfrak{F}_{D,fin}$ is dense in $\mathfrak{F}_D$, inequality
\eqref{n87} holds for all $\Phi\in\mathfrak{F}_D$ and all triples
$(q,r,s)$. Therefore, we have
\begin{equation}\nonumber
\|B_+\Psi^{(q,r,s)}\|^2 \leq \|G_+\|^2 \| (N_D+1)^\frac12
\Psi^{(q,r,s)}\|^2
\end{equation}
which yields
\begin{equation}\label{n90}
\|B_+\Psi\| \leq \|G_+\|~\| (N_D+1)^\frac12 \Psi\|
\end{equation}
for every $\Psi\in\mathfrak{F}_{D,fin}$. Since
$\mathfrak{F}_{D,fin}$ is a core for $(N_D+1)^\frac12$,
$\mathfrak{D}(B_+) \supset \mathfrak{D}((N_D+1)^\frac12 )$ and
\eqref{n90} is still true for every
$\Psi\in\mathfrak{D}((N_D+1)^\frac12)$.

The proof of \eqref{eq:31.bis} and \eqref{eq:31.ter} is similar to
the above one, if we use in addition that for all $\gamma, n$,
$\|b^*_{\gamma,n}\| =1$.  This concludes the proof of
Proposition~\ref{prop2.5}.
\end{proof}
\begin{rema}\label{nrem1}
Inequalities \eqref{eq:31} and \eqref{eq:32} are the best
estimates that we can get. Indeed, set
\begin{equation}\nonumber
\Phi^{(l,m,n)} = \prod_{i=1}^\ell b_d^*(f_i)\otimes \prod_{j=1}^m
b_+^*(g_j) \otimes \prod_{k=1}^n b_-^*(h_k) \Omega
\end{equation}
we have
\begin{equation}\label{n92}
\begin{split}
 \lefteqn{B_+\Phi^{(l,m,n)} = \sum_{\alpha=1}^m \langle G_+,
 g_\alpha\otimes g_\alpha \rangle \Phi^{(l,m,n)}} &  \\
 & +\! \sum_{\alpha=1}^m (-1)^{\alpha+1}\!\! \sum_{\beta =m+1}^\infty
 \!\!\langle G_+, g_\beta\otimes g_\alpha\rangle \prod_{i=1}^\ell
 b_d^*(f_i) \otimes b_+^*(g_\beta)\!\! \prod_{\buildrel{j\in\{1,\ldots,m\}}
 \over{j\neq\alpha}}\!\!b_+^*(g_j) \otimes \prod_{k=1}^n
 b_-^*(h_k)\Omega .
\end{split}
\end{equation}
Two different vectors in the right hand side of \eqref{n92} are
orthogonal. Therefore, we get
\begin{equation}\nonumber
\|B_+ \Phi^{(\ell,m,n)}\|^2 = \left| \sum_{\alpha=1}^m\langle
G_+,g_\alpha \otimes g_\alpha\rangle\right|^2 + \sum_{\alpha=1}^m
\sum_{\beta=m+1}^\infty |\langle G_+, g_\beta\otimes
g_\alpha\rangle |^2
\end{equation}
from which we deduce
\begin{equation}\nonumber
\|B_+ \Phi^{(\ell,m,n)}\|^2 \leq m
\|G_+\|^2\|\Phi^{(\ell,m,n)}\|^2 \leq \|G_+\|^2
\|N_D^\frac12\Phi^{(\ell,m,n)}\|^2 .
\end{equation}
\end{rema}
%%%%%%%%%%%%%%%%%%%%%%%% fin CORRECTIONS 1 %%%%%%%%%%%%%%%%%%%%%%%

We now have the following two lemmata whose proof is easy.

\begin{lem}\label{lem2.6}
Let $A$ (resp. $B$) be a positive self-adjoint operator in the
Hilbert space ${\mathfrak H}_1$ (resp. ${\mathfrak H}_2$) with
domain $\D(A)$ (resp. $\D(B)$).

Then the operator $H_0 = A\otimes \1 + \1 \otimes B$ with domain $\D(A\otimes
\1) \cap \D(\1 \otimes B)$ is self-adjoint in ${\mathfrak H}_1 \otimes
{\mathfrak H}_2$ and we have
\begin{eqnarray*}
\|(A \otimes \1)\, \psi\| & \leq & \|H_0\, \psi\| ,\\ \noalign{\vskip 8pt}
\|(\1 \otimes B)\, \psi\| & \leq & \|H_0\, \psi\| ,
\end{eqnarray*}
for every $\psi \in \D(A\otimes \1) \cap \D(\1 \otimes B)$.
\end{lem}

\begin{lem}\label{lem2.7}
Let $E_0={\rm inf}_{(\gamma,n)\in {\mathcal D}_d}E_{\gamma,n}>0$. We have
 $$
  E_0 \|N_D\,
  \Psi\| \leq \|d\Gamma(H_D)\, \Psi\|
 $$
for any $\Psi \in \D(d\Gamma(H_D)).$
\end{lem}

Applying Lemma~\ref{lem2.6} to $A = d\Gamma(H_D)$ and  $B=H_{ph}$ we get, from
Lemma~\ref{lem2.7}, the following result

\begin{prop}\label{prop2.8}
For any $\Psi \in \D(d\Gamma(H_D) \otimes \1) \cap \D(\1 \otimes H_{ph})$, we
have
\begin{eqnarray*}
 E_0 \|\left(N_D \otimes \1\right) \Psi\| &\leq & \|H_0 \Psi\|\\
 \noalign{\vskip 8pt}
 \|\left(\1 \otimes H_{ph}\right)\Psi \| &\leq & \|H_0 \Psi\|
\end{eqnarray*}
where $H_0 = d\Gamma(H_D) \otimes \1 + \1 \otimes H_{ph}$
\end{prop}

\subsection{Proof of Theorem~\ref{thm1}}
The proof of Theorem~\ref{thm1} is achieved by showing that
$H_I^{(1)}$ and $H_I^{(2)}\otimes\1$ are relatively $H_0$-bounded.

We first treat $H_I^{(1)}$. The result follows from
Lemma~\ref{lem2.9} and Corollary~\ref{corr2.10}. Then we study all
the quartic terms that appear in $H_I^{2}$, and the results are
collected in Corollary~\ref{corr-quartic}. In both cases, details
of proofs are given in the Appendix.

%%%%%%%%%%%%%%%%%%%%%%%%%%%%%%%%%%%%%%%%%%%%%%%%%%%%%%%%%%%%
%%%%%%%% RELATIVE BOUNDS FOR H_I^{(1)} %%%%%%%%%%%%%%%%%%%%%
%%%%%%%%%%%%%%%%%%%%%%%%%%%%%%%%%%%%%%%%%%%%%%%%%%%%%%%%%%%%
The interaction between electrons, positrons and transversal
photons can be written in the following form (See
Section~\ref{S2.4})
\begin{equation}\label{eq:39}
H_I^{(1)} = \sum_{i=1}^6 \sum_{\mu =1,2} \int \d^3 k(v_ i^\mu(k) \otimes
a_\mu^*(k) + {v_i^\mu}^*(k) \otimes a_\mu(k))
\end{equation}
where
\begin{equation}\nonumber
\begin{split}
v_1^\mu(k) &= \sum_{\gamma,\gamma',n,\ell}
G_{d,\gamma,\gamma',n,\ell}^\mu(k) b_{\gamma,n}^* b_{\gamma',l}\
,\\ v_2^\mu(k) &= \sum_{\gamma,\gamma',n} \int \d p\,
G_{d,+,\gamma,\gamma',n}^\mu(p;k) (b^*_{\gamma,n} b_{\gamma',+}(p)
+ b^*_{\gamma,+}(p)b_{\gamma',n}) , \\ v_3^\mu(k) &=
\sum_{\gamma,\gamma',n} \int \d p\,
G_{d,-,\gamma,\gamma',n}^\mu(p;k)
(b^*_{\gamma,n}b^*_{\gamma',-}(p) + b_{\gamma,-}(p)b_{\gamma',n})\
,\\ v_4^\mu(k) &= \sum_{\gamma,\gamma'} \int\int \d p\,\d p'
G_{+,-,\gamma,\gamma'}^\mu(p,p';k) (b_{\gamma,+}^*(p)
b_{\gamma',-}^* (p') +b_{\gamma,-}(p) b_{\gamma',+}(p')) ,\\
v_5^\mu(k) &= \sum_{\gamma,\gamma'} \int\int \d p\,\d p'
G_{+,+,\gamma,\gamma'}^\mu(p,p';k) b_{\gamma,+}^*(p) b_{\gamma',+}
(p') ,
\\
v_6^\mu(k) &= \sum_{\gamma,\gamma'} \int\int \d p\,\d p'
G_{-,-,\gamma,\gamma'}^\mu(p,p';k) b_{\gamma,-}^*(p) b_{\gamma',-} (p') .
\end{split}
\end{equation}
It follows from Proposition~\ref{nprop3.4},
Proposition~\ref{prop2.5} that $v_i^\mu(k)$, $i \not= 5,6$,
$v_j^\mu(k) (N_D+1)^{-1/2}$, $(N_D+1)^{-1/2} v_j^\mu(k)$, $j =
5,6$ are bounded from ${\mathfrak F}_D$ into ${\mathfrak F}_D$.

For $\beta = 0,1$, $\mu = 1,2$ and $i \in \{1,2,\ldots, 6\}$, we set
\begin{equation}
\begin{split}\label{eq:41}
a_{\beta,i}^\mu &= \left( \int_{\er^3} \omega(k)^{-\beta} \|v_i^\mu(k)\|^2
\d^3k \right)^{1/2}, \qquad i \not= 5,6\\ a_{\beta,i}^\mu &= \left(
\displaystyle\int_{\er^3} \omega(k)^{-\beta} \|v_i^\mu(k) (N_D+1)^{-1/2}\|^2
\d^3k \right)^{1/2}, \qquad i = 5,6.
\end{split}
\end{equation}
\begin{equation}\label{eq:42}
\begin{split}
b_{i}^\mu &= \left( \int \omega(k)^{-1} \|v_i^\mu(k)\|^2 \d^3k \right)^{1/2},
\qquad i \not= 5,6\\
 b_{i}^\mu &= \left( \int \omega(k)^{-1}
\|(N_D+1)^{-1/2}\, v_i^\mu(k)\|^2 \d^3k \right)^{1/2}, \qquad i = 5,6.
\end{split}
\end{equation}
\begin{lem}\label{lem2.9}
For every $\Psi \in \D(H_0)$, we have
\begin{equation}\label{eq:43}
\left\Vert \int \d^3k\, v_i^\mu(k)^* \otimes a_\mu(k)\, \Psi \right\Vert \leq
\frac{b_i^\mu}{{\sqrt{E_0}}} \|( H_0 + m_0 c^2 ) \Psi\|,
\end{equation}
\begin{equation}\label{eq:44}
\begin{split}
\left\Vert \int \d^3k\, v_i^\mu(k) \otimes a_\mu^*(k)\, \Psi
\right\Vert^2 &\leq \left(\frac{(a_{1,i}^\mu)^2}{E_0} +
\varepsilon \frac{(a_{0,i}^\mu)^2}{E_0^2}\right)
\|(H_0+m_0c^2)\Psi\|^2\\ &+ \frac{(a_{0,i}^\mu)^2}{4\varepsilon}
\|\Psi\|^2
\end{split}
\end{equation}
for every $\varepsilon > 0$.
\end{lem}
\begin{proof}
In the appendix, we prove that, for every $\Psi \in \D(H_0)$ and every
$\varepsilon > 0$, we have
\begin{equation}\label{eq:45}
\left\Vert \int \d^3k\, v_i^\mu(k)^* \otimes a_\mu(k)\, \Psi \right\Vert \leq
b_i^\mu \, \|(N_D+1)^{1/2} \otimes H_{ph}^{1/2} \Psi\|,
\end{equation}
\begin{eqnarray}\label{eq:46}
&&\left\Vert \int \d^3k\, v_i^\mu(k) \otimes a_\mu^*(k)\, \Psi
\right\Vert^2 \leq (a_{1,i}^\mu)^2 \|(N_D+1)^{1/2} \otimes
H_{ph}^{1/2} \Psi\|^2 \nonumber\\ \noalign{\vskip 8pt}
&&\qquad\qquad\qquad + (a_{0,i}^\mu)^2 \left[ \varepsilon
\|(N_D+1)\otimes\1\, \Psi\|^2 +
 \frac{1}{4\varepsilon} \|\Psi\|^2\right],
\end{eqnarray}
which together with Lemma~\ref{lem2.6} and Lemma~\ref{lem2.7} give
\eqref{eq:43} and \eqref{eq:44} by noting that $E_0<m_0c^2.$
\end{proof}

Finally, from \eqref{eq:39}, \eqref{eq:43} and \eqref{eq:44}, we get

\begin{corr}\label{corr2.10}
The operator $H_I^{(1)}$ is $(H_0+m_0 c^2)$-bounded, with relative
bound
 $$
  C'_1 = \sum_{\mu =1,2} \sum_{i=1}^6 \frac{a_{1,i}^\mu +
  b_{i}^\mu}{{\sqrt{E_0}}}  .
 $$
\end{corr}

%%%%%%%%%%%%%%%%%%%%%%%%%%%%%%%%%%%%%%%%%%%%%%%%%%%%%%%%%%%%
%%%%%%%% RELATIVE BOUNDS FOR H_I^{(2)} %%%%%%%%%%%%%%%%%%%%%
%%%%%%%%%%%%%%%%%%%%%%%%%%%%%%%%%%%%%%%%%%%%%%%%%%%%%%%%%%%%

%%%% debut corrections 2 %%%%%%%%%%%%

We now treat the term $H_I^{(2)}$. Again in the case of electrons, we
restrict ourselves to the electrons in the continuous spectrum.  The
case of electrons in the discrete spectrum will be simpler to deal
with.

For $F^{(1)}\in L^2$, the following formal operator
 $$
  \int\d\xi_1\,\d\xi_2\,\d\xi_3\,\d\xi_4
  F^{(1)}(\xi_1,\xi_2,\xi_3,\xi_4) b_+^*(\xi_1) b_-^*(\xi_2)
  b_+(\xi_3) b_-(\xi_4)
  $$
is defined as a quadratic form on $\mathfrak{F}_{D,fin}\times
\mathfrak{F}_{D,fin}$ in the following way
\begin{equation}\nonumber
 \int \d\xi_1 \d\xi_2 \d\xi_3 \d\xi_4 \langle b_-(\xi_2)
 b_+(\xi_1) \Psi, b_+(\xi_3) b_-(\xi_4)\Phi\rangle
\end{equation}
As for the quadratic terms, one can show that there exists an
operator, denoted by $C^{(1)}$ such that $C^{(1)}$ is the unique
operator in $\mathfrak{F}_D$ such that
$\mathfrak{F}_{D,fin}\subset \mathfrak{D}(C^{(1)})$ is a core for
$C^{(1)}$ and
\begin{equation}\nonumber
C^{(1)} = \int\d\xi_1\d\xi_2\d\xi_3\d\xi_4 F^{(1)}(\xi_1,
\xi_2,\xi_3,\xi_4) b_+^*(\xi_1) b_-^*(\xi_2) b_+(\xi_3) b_-(\xi_4)
\end{equation}
and
\begin{equation}\nonumber
(C^{(1)})^* = \int\d\xi_1\d\xi_2\d\xi_3\d\xi_4
\overline{F^{(1)}(\xi_3, \xi_4,\xi_1,\xi_2)} b_+^*(\xi_1)
b_-^*(\xi_2) b_+(\xi_3) b_-(\xi_4)
\end{equation}
as quadratic forms on $\mathfrak{F}_{D,fin}\times
\mathfrak{F}_{D,fin}$. In the same way, for $F_\epsilon^{(2)}$,
$F_{\epsilon,\epsilon'}^{(3)}$ and $F^{(4)}$
in $L^2$, we define the following operators as quadratic forms on
$\mathfrak{F}_{D,fin}\times\mathfrak{F}_{D,fin}$
\begin{equation}\nonumber
C^{(2)}_{\epsilon} = \int\d\xi_1\d\xi_2\d\xi_3\d\xi_4
{F^{(2)}_\epsilon(\xi_1, \xi_2,\xi_3,\xi_4)} b_\epsilon^*(\xi_1)
b_\epsilon^*(\xi_2) b_\epsilon(\xi_3) b_\epsilon(\xi_4) ,
\end{equation}
\begin{equation}\nonumber
(C^{(2)}_{\epsilon})^* = \int\d\xi_1\d\xi_2\d\xi_3\d\xi_4
\overline{F^{(2)}_\epsilon(\xi_4, \xi_3,\xi_2,\xi_1)}
b_\epsilon^*(\xi_1) b_\epsilon^*(\xi_2) b_\epsilon(\xi_3)
b_\epsilon(\xi_4) ,
\end{equation}
\begin{equation}\nonumber
C^{(3)}_{\epsilon,\epsilon'} = \int\d\xi_1\d\xi_2\d\xi_3\d\xi_4
{F^{(3)}_{\epsilon,\epsilon'}(\xi_1, \xi_2,\xi_3,\xi_4)}
b_\epsilon^*(\xi_1) b_\epsilon(\xi_2) b_{\epsilon'}(\xi_3)
b_\epsilon(\xi_4) ,
\end{equation}
\begin{equation}\nonumber
(C^{(3)}_{\epsilon,\epsilon'})^* = -
\int\d\xi_1\d\xi_2\d\xi_3\d\xi_4
\overline{F^{(3)}_{\epsilon,\epsilon'}(\xi_4, \xi_2,\xi_3,\xi_1)}
b_\epsilon^*(\xi_1) b_\epsilon^*(\xi_2) b_{\epsilon'}^*(\xi_3)
b_\epsilon(\xi_4) ,
\end{equation}
\begin{equation}\nonumber
C^{(4)} = \int\d\xi_1\d\xi_2\d\xi_3\d\xi_4
{F^{(4)}(\xi_1, \xi_2,\xi_3,\xi_4)}
b_+(\xi_1) b_+(\xi_2) b_-(\xi_3)
b_-(\xi_4) ,
\end{equation}
\begin{equation}\nonumber
(C^{(4)})^* =
\int\d\xi_1\d\xi_2\d\xi_3\d\xi_4
\overline{F^{(4)}(\xi_1, \xi_2,\xi_3,\xi_4)}
b_+^*(\xi_1) b_+^*(\xi_2) b_-^*(\xi_3)
b_-^*(\xi_4) ,
\end{equation}
We now have the following proposition
\begin{prop}\label{newquarticbound}$\ $

\begin{itemize}
\item[{(i)}] For $F^{(1)}\in L^2$ we have
 $$
  \| C^{(1)} (N_D +1)^{-1} \|\leq \frac{1}{\sqrt2} \|F^{(1)}\|
 $$
and
 $$
  \| (C^{(1)})^* (N_D +1)^{-1} \|\leq \frac{1}{\sqrt2} \|F^{(1)}\|
 $$

\item[{(ii)}] For $F^{(2)}_\epsilon\in L^2$ we have
 $$
  \| C^{(2)}_{\epsilon} (N_D +1)^{-1} \| \leq \|
  F_{\epsilon}^{(2),a}\|
 $$
and
 $$
  \| (C^{(2)}_{\epsilon})^* (N_D +1)^{-1} \| \leq \|
  F_{\epsilon}^{(2),a}\|
 $$
where $F_{\epsilon}^{(2),a}(\xi_1,\xi_2,\xi_3,\xi_4) =
F_{\epsilon}^{(2)}(\xi_1,\xi_2,\xi_3,\xi_4) -
F_{\epsilon}^{(2)}(\xi_1,\xi_2,\xi_4,\xi_3)$.

\item [{(iii)}] For $F^{(3)}_{\epsilon,\epsilon'}\in L^2$ we have
 $$
  \| C^{(3)}_{\epsilon,\epsilon'} (N_D +1)^{-\frac12} \| \leq \|
  F_{\epsilon,\epsilon'}^{(3),a}\|
 $$
and
 $$
  \| (C^{(3)}_{\epsilon,\epsilon'})^* (N_D +1)^{-\frac12} \| \leq \|
  F_{\epsilon,\epsilon'}^{(3),a}\|
 $$
where $F_{\epsilon,\epsilon'}^{(3),a}(\xi_1,\xi_2,\xi_3,\xi_4) =
F_{\epsilon,\epsilon'}^{(3)}(\xi_1,\xi_2,\xi_3,\xi_4) -
F_{\epsilon,\epsilon'}^{(3)}(\xi_1,\xi_4,\xi_3,\xi_2)$.

\item[{(iv)}] For $F^{(4)}\in L^2$ we have
 $$
  \| C^{(4)} \| \leq \|
  F^{(4),aa}\|
 $$
and
 $$
  \| (C^{(4)})^*  \| \leq \|
  F^{(4),aa}\|
 $$
\begin{sloppypar}\noindent where
$F^{(4),aa}(\xi_1,\xi_2,\xi_3,\xi_4) =
F^{(4)}(\xi_1,\xi_2,\xi_3,\xi_4) -
F^{(4)}(\xi_1,\xi_2,\xi_4,\xi_3) -
F^{(4)}(\xi_2,\xi_1,\xi_3,\xi_4) +
F^{(4)}(\xi_2,\xi_1,\xi_4,\xi_3)$.
\end{sloppypar}
\end{itemize}
\end{prop}
\begin{rema}
The estimates of Proposition~\ref{newquarticbound} are satisfied
both for electrons in the continuous spectrum and the discrete one
with appropriate $L^2$-norms.
\end{rema}
\begin{proof}
As mentioned above, for the proof in the case of electrons, we
only consider electrons in the continuous spectrum. The proof for
electrons in the discrete spectrum is similar and simpler. We
mimick the proof of Proposition~\ref{prop2.5} by using
Proposition~\ref{nprop3.4}. We first prove (i). Let $\Psi =
(\Psi^{(q,r,s)})$ and $\Phi = (\Phi^{(q,r,s)})$ be two vectors of
$\mathfrak{F}_{D,fin}$. We have
\begin{equation}\nonumber
\begin{split}
 \lefteqn{\langle \Phi^{(q',r',s')}, C^{(1)} \Psi^{(q,r,s)}\rangle
 =\delta_{q,q'} \delta_{r,r'} \delta_{s,s'}} & \\
 & \times\int
 \d\xi_1\d\xi_2\d\xi_3\d\xi_4 F^{(1)}(\xi_1,\xi_2,\xi_3,\xi_4)
 \langle b_-(\xi_2)b_+(\xi_1) \Phi^{(q,r,s)}, b_+(\xi_3)
 b_-(\xi_4) \Psi^{(q,r,s)}\rangle
\end{split}
\end{equation}
Thus $C^{(1)}\Psi^{(q,r,s)}\in \mathfrak{F}_a^{(q,r,s)}$ for every
triple $(q,r,s)$. By the Fubini Theorem we get
\begin{equation}\nonumber
 \begin{split}
 \lefteqn{|\langle \Phi^{(q',r',s')}, C^{(1)}
 \Psi^{(q,r,s)}\rangle|} & \\
 & = \left|\int\!\! \left\langle b_-(\xi_2) b_+(\xi_1) \Psi^{(q,r,s)},
 \!\!\int\!\! F^{(1)} (\xi_1,\xi_2,\xi_3,\xi_4) b_+(\xi_3) b_-(\xi_4)
 \Psi^{(q,r,s)} \d\xi_3 \d\xi_4\right\rangle \d\xi_1\d\xi_2 \right|
\end{split}
\end{equation}
By the Cauchy-Schwarz inequality and
Proposition~\ref{nprop3.4}(ii) we get
\begin{equation}\nonumber
 \begin{split}
  \lefteqn{|\langle \Phi^{(q,r,s)}, C^{(1)} \Psi^{(q,r,s)}\rangle|^2
  \leq} & \\
  & \left(\int \|b_-(\xi_2) b_+(\xi_1) \Psi^{(q,r,s)}\| \left(
  \int | F^{(1)} (\xi_1,\xi_2,\xi_3,\xi_4)|^2
  \d\xi_3 \d\xi_4\right)^{\frac12}\!\! \d\xi_1\d\xi_2\! \right)^2
  \!\!\|\Psi^{(q,r,s)}\|^2
 \end{split}
\end{equation}
Again by the Cauchy-Schwarz inequality and the definitions of
$b_+(\xi)$ and $b_-(\xi)$ we finally get
 \begin{eqnarray*}
  |\langle \Phi^{(q,r,s)}, C^{(1)}\Psi^{(q,r,s)}\rangle|^2 & \leq &
  r s
  \|F^{(1)}\|^2 \|\Phi^{(q,r,s)}\|^2 \|\Psi^{(q,r,s)}\|^2 \\& \leq &
  \frac12 \|F^{(1)}\|^2 \|\Phi^{(q,r,s)}\|^2
  \|(N_D+1)^2\Psi^{(q,r,s)}\|^2 ,
 \end{eqnarray*}
which yields
 $$
  \|C^{(1)} (N_D+1)^{-1} \| \leq \frac{1}{\sqrt 2} \|F^{(1)}\|_2\
  .
 $$
The estimate for $(C^{(1)})^*$ follows directly from above and
from
 $$
  \langle (C^{(1)})^* \Phi^{(q,r,s)}, \Psi^{(q,r,s)}\rangle =
  \langle (\Phi^{(q,r,s)}, C^{(1)} \Psi^{(q,r,s)}\rangle
 $$
Let us now prove (ii). We consider only the case of $C^{(2)}_+$,
since the proof for $C^{(2)}_-$ is the same. We have
\begin{equation}\nonumber
\begin{split}
\lefteqn{ \langle \phi^{(q',r',s')}, C_+^{(2)} \Psi^{(q,r,s)}
\rangle = \delta_{qq'} \delta_{rr'} \delta_{ss'}} &\\
 & \times \int
\d\xi_1\d\xi_2\d\xi_3\d\xi_4 F_+^{(2)}(\xi_1, \xi_2, \xi_3,\xi_4)
\langle b_+(\xi_2) b_+(\xi_1) \Phi^{(q,r,s)}, b_+(\xi_3)
b_+(\xi_4) \Psi^{(q,r,s)}\rangle
\end{split}
\end{equation}
Thus $C^{(2)}_+\Psi^{(q,r,s)}\in\mathfrak{F}_a^{(q,r,s)}$ for
every triple $(q,r,s)$ and
\begin{equation}\nonumber
\begin{split}
\lefteqn{ |\langle \Phi^{(q,r,s)}, C_+^{(2)}
\Psi^{(q,r,s)}\rangle|^2 }&
\\
 = &
\left|\int\left\langle b_+(\xi_2) b_+(\xi_1) \Phi^{(q,r,s)}, \int
F_+^{(2)}(\xi_1, \xi_2, \xi_3, \xi_4) b_+(\xi_3) b_+(\xi_4)
\Psi^{(q,r,s)} \right\rangle \d\xi_1\d\xi_2\right|^2
\end{split}
\end{equation}
By the Cauchy-Schwarz inequality and Proposition~\ref{nprop3.4}(i)
we obtain
\begin{equation}\nonumber
\begin{split}
\lefteqn{ |\langle \Phi^{(q,r,s)}, C_+^{(2)}
\Psi^{(q,r,s)}\rangle|^2 }&
\\ &\leq\!\!
\left(\! \int\! \| b_+(\xi_2) b_+(\xi_1) \Phi^{(q,r,s)}\|
\!\left(\int \!\!|F_+^{(2),a}(\xi_1, \xi_2, \xi_3, \xi_4)|^2
\d\xi_3\d\xi_4\right)^\frac12 \!\!\!\d\xi_1\d\xi_2\!\!\right)^2
\!\!\|\Psi^{(q,r,s)}\|^2
\end{split}
\end{equation}
Applying again Cauchy-Schwarz inequality and the definition of
$b_+(\xi)$ we get
\begin{equation}\nonumber
\begin{split}
{|\langle \Phi^{(q,r,s)}, C_+^{(2)} \Psi^{(q,r,s)}\rangle|^2 } &
\leq\! r(r-1) \|F_+^{(2),a}\|^2 \|\Phi^{(q,r,s)}\|^2 \|
\Psi^{(q,r,s)}\|^2 \\ & \leq \|F_+^{(2),a}\|^2
\|\Phi^{(q,r,s)}\|^2 \|(N_D+1) \Psi^{(q,r,s)}\|^2 .
\end{split}
\end{equation}
We conclude this item as in the proof of
Proposition~\ref{prop2.5}. We thus obtain
\begin{equation}\nonumber
 \| C_+^{(2)}(N_D + 1)^{-1} \| \leq \| F_\epsilon^{(2),a} \| .
\end{equation}
The proof for $C_-^{(2)}$, $(C_+^{(2)})^*$ and $(C_-^{(2)})^*$ is
similar.

We now prove (iii). We have
\begin{equation}\nonumber
\begin{split}
\lefteqn{\langle\Phi^{(q',r',s')},C_{+-}^{(3)}\Psi^{(q,r,s)}\rangle
= \delta_{q',q}\delta_{r',r-1}\delta_{s', s-1}}& \\ &\times
\int\d\xi_1\d\xi_2\d\xi_3\d\xi_4 F_{+-}^{(3)}
(\xi_1,\xi_2,\xi_3,\xi_4) \langle b_+(\xi_1)\Phi^{(q',r',s')},
b_+(\xi_2)b_-(\xi_3)b_+(\xi_4)\Psi^{(q,r,s)}\rangle
\end{split}
\end{equation}
Thus $C_{+-}^{(3)}\Psi^{(q,r,s)}\in \mathfrak{F}_a^{q,r-1,s-1)}$
for $r>1$ and $s>1$ and $C_{+-}^{(3)}\Psi^{(q,r,s)}=0$ for $r\leq
1$ or $s\leq 1$. We then have
\begin{equation}\nonumber
\begin{split}
\lefteqn{| \langle \Phi^{(q,r-1,s-1)}, C_{+-}^{(3)}
\Psi^{(q,r,s)}\rangle|^2} & \\ &  = \left| \int \left\langle
b_+(\xi_1)\Phi^{(q,r-1,s-1)}, \int F_{+-}^{(3)} (\xi_1,
\xi_2,\xi_3,\xi_4) b_+(\xi_2) b_-(\xi_3) b_+(\xi_4)
\Psi^{(q,r,s)}\right\rangle \right|^2
\end{split}
\end{equation}
Using Cauchy-Schwarz inequality  and
Proposition~\ref{nprop3.4}(iii) we obtain
\begin{equation}\nonumber
\begin{split}
\lefteqn{| \langle \Phi^{(q,r-1,s-1)}, C_{+-}^{(3)}
\Psi^{(q,r,s)}\rangle|^2} & \\ &\!  \leq\! \left(\! \int\! \|
b_+(\xi_1)\Phi^{(q,r-1,s-1)}\| \left(\int\! |F_{+-}^{(3),a}
(\xi_1, \xi_2,\xi_3,\xi_4)|^2 \d\xi_2 \d\xi_3
\d\xi_4\right)^{\frac12}\!\!\d\xi_1\right)^2\!\!
\|\Psi^{(q,r,s)}\|^2
\end{split}
\end{equation}
Thus, using Cauchy-Schwarz inequality and the definition of
$b_+(\xi)$ we get
\begin{equation}\nonumber
\begin{split}
 | \langle \Phi^{(q,r-1,s-1)}, C_{+-}^{(3)}
 \Psi^{(q,r,s)}\rangle|^2 &   \leq
 (r-1) \|F_{+-}^{(3),a}\|^2 \|\Phi^{(q,r-1,s-1)}\|^2
 \|\Psi^{(q,r,s)}\|^2 \\
 & \leq \|F_{+-}^{(3),a}\|^2 \|\Phi^{(q,r-1,s-1)}\|^2
 \|(N_D + 1)^{\frac12}\Psi^{(q,r,s)}\|^2
\end{split}
\end{equation}
We conclude as in the proof of Proposition~\ref{prop2.5} that
\begin{equation}\nonumber
\|C_{+-}^{(3)} (N_D + 1)^{-\frac12}\| \leq \|F_{+-}^{(3),a}\| .
\end{equation}
The proofs for $C_{\epsilon\epsilon'}^{(3)}$ and
$(C_{\epsilon\epsilon'}^{(3)})^*$ are very similar.

Finally, by Proposition~\ref{nprop3.4}(iv) we get immediately
\begin{equation}\nonumber
 \|C^{(4)} \| \leq \|F^{(4),aa}\|\
\end{equation}
and
\begin{equation}\nonumber
 \|(C^{(4)})^* \| \leq \|F^{(4),aa}\|\
\end{equation}

This concludes the proof of Proposition~\ref{newquarticbound}.
\end{proof}
\begin{rema}
 One can show, as in Remark~\ref{nrem1} that the estimates (i),
 (ii), (iii) and (iv) are the best one can get.
\end{rema}

 As a consequence of Lemmas~\ref{lem2.6}, ~\ref{lem2.7} and
Propositions~\ref{nprop3.4}, \ref{prop2.5} and \ref{newquarticbound}
we obtain
\begin{corr}\label{corr-quartic}
For any $\Psi\in \D(N_D)$ we have
\begin{equation}\nonumber
\begin{split}
  \lefteqn{\|H_I^{(2)}\Psi\|  \leq \left( \frac{1}{\sqrt 2}\| F^{(1)}\| +
  \| F_+^{(2),a}\| + \|F_-^{(2),a}\|\right) \| (N_D +
  1)\Psi\|} & \\
  &  + \left[ 2 \left(\|F_{+-}^{(3),a}\| + \|F_{-+}^{(3),a}\| \right)
  + \|F_+^{(5)}\| + \|F_-^{(5)}\| \right] \|(N_D + 1 )^{\frac12} \Psi \| \\
  &  + 2 \left[  \|F^{(4)}\|
  +   \|F_{+-}^{(6)}\| + \|F_{-+}^{(6)}\|  \right]
  \|\Psi\| .
\end{split}
\end{equation}
\end{corr}

\noindent{\it Proof of Theorem~\ref{thm1}.} Supposing that $C_0 <
\infty$, $C_1<\infty$ and choosing $g_1$ and $g_2$ such that
$\frac{|g_1|}{\sqrt{E_0}}C_1 + \frac{|g_2|}{E_0}
(\frac{1}{\sqrt{2}} \|F^{(1)}\| + \|F^{(2),a}_+\| +
\|F^{(2),a}_-\| )<1$, Theorem~\ref{thm1} follows from
Proposition~\ref{prop2.5}, Corollary~\ref{corr2.10},
Corollary~\ref{corr-quartic} and Theorem V.4.3 of \cite{Kato}.

\section{Proof of Theorem~\ref{thm2}}\label{section4}
Throughout this section we assume that the assumptions of Theorem~\ref{thm2}
are satisfied. Let us introduce an infrared cutoff in the first part of the
interaction

Let  $H_{I,m}^{(1)}$ be the operator obtained from \eqref{eq:39} by
substituting
 $$
   v_{i,m}^\mu(k)  := 1_{\{k;\omega(k) \geq m\}} v_i^\mu(k) ,\ m >
   0 ,
 $$
 for $v_i^\mu(k)$. We then define
\begin{equation}\nonumber
H_m : = H_0 + g_1  H_{I,m}^{(1)} + g_2 H_I^{(2)}\otimes\1.
\end{equation}

Theorem~\ref{thm2} will be a simple consequence of the following result.

\begin{theo}\label{thm3}
There exists $g_0>0$ such that for every $(g_1,\, g_2)$ satisfying
$|g_1| + |g_2| \leq g_0$, the following properties hold.
\begin{itemize}
\item[(i)]  For every $\Psi \in \D(H_0)$,
$\displaystyle H_m\, \Psi \mathop{\longrightarrow}_{m\mapsto
0} H\Psi$.
\item[(ii)] For every $m \in (0,1)$, $H_m$ has a normalized ground state
$\Phi_m$.
\item[(iii)] Fix $\lambda$ in $(E_0, m_0\, c^2)$.
For every $m > 0$, we have
\begin{equation}\label{eq:48}
\la\Phi_m, P_{(-\infty,\lambda]}(d\Gamma(H_D)) \otimes P_{\Omega_{ph}}
\Phi_m\ra  \geq 1 - \delta_{g_1,\, g_2} ,
\end{equation}
where $\delta_{g_1,\,g_2}$ tends to zero when $|g_1| + |g_2|$
tends to zero and $\delta_{g_1,g_2} <1$ for $|g_1| + |g_2| \leq
g_0$.
\end{itemize}
\end{theo}
Theorem~\ref{thm2} is easily deduced from Theorem~\ref{thm3} as
follows: Let $H_m$ and $\Phi_m$ be the same as in
Theorem~\ref{thm3}. Since $\|\Phi_m\| = 1$, there exists a
subsequence $(\Phi_{m_k})$ of $(\Phi_m)$ such that $w-\lim
\Phi_{m_k} = \Phi$.

On the other hand, since $P_{(-\infty,\lambda]}(d\Gamma(H_D))
\otimes P_{\Omega_{ph}}$ is finite rank for $\lambda \in (E_0, m_0
c^2)$, it follows from \eqref{eq:48} that
 $$
 \la\Phi, P_{(-\infty, \lambda]}(d\Gamma(H_D)) \otimes P_{\Omega_{ph}}\Phi\ra
 \geq 1-\delta_{g_1,\,g_2},
 $$
which implies that $\Phi \not= 0$.

Now, in order to conclude the proof of Theorem~\ref{thm2}, it suffices to apply
the following well known result \cite{ref12}

\begin{lem}\label{lem3.2}
Let $T_n,\ n\geq1,$ and  $T$ be self-adjoint operators on a
Hilbert space $\mathfrak{K}$ having a common core $\mathfrak{D}$
such that, for all $\Phi\in\mathfrak{D}$, $T_n\Phi\rightarrow
T\Phi$ as $n\rightarrow\infty$. We assume that,  for every
$n\geq1,\ T_n$ has a normalized ground state $\Phi_n$ with ground
state energy $E_n$ such that $\lim_{n\rightarrow \infty} E_n =E$
and $\omega - \lim_{n\mapsto +\infty} \Phi_n = \Phi \not= 0$. Then
$\Phi$ is a ground state of $T$ with ground state energy $E$.
\end{lem}

\bigskip

Theorem~\ref{thm3} will be proved in several steps. We first prove (i). We then
prove (iii). Finally, by the periodic approximation method, we will prove the
existence of a ground state for $H_m$, $m > 0$.

\subsection{Proof of (i) of Theorem~\ref{thm3}}
\begin{proof}
For $m > 0$, set
 $$
   \tilde v_i^{\mu,m}(k) = 1_{\{k;\omega(k) < m\}}(k) \,
   v_i^\mu(k).
 $$ We denote by $\tilde a_{\beta,j}^{\mu,m}$ (resp. $\tilde
b_j^{\mu,m}$) the expression \eqref{eq:41} and \eqref{eq:42} that we
obtain by substituting $\tilde v_j^{\mu,m}(k)$ for $v_j^\mu(k)$.

Since $H-H_m = g_1 (H_I^{(1)} - H_{I,m}^{(1)})$, it follows from
Lemma~\ref{lem2.9} that
\begin{equation}\nonumber
  \|(H-H_m) \Phi\| \leq |g_1|\!\! \sum_{\mu=1,2}\, \sum_{j=0}^6\!\left(
  \left(\frac{\tilde a_{1,j}^{\mu,m} + \tilde b_j^{\mu,m}}{\sqrt{E_0}}
  + \frac{\tilde a_{0,j}^{\mu,m}}{E_0}\right)\!\!
  \, \| (H_0+m_0c^2)\Phi\| + \frac{\tilde a_{0,j}^{\mu,m}}
  {4} \|\Phi\|\right)
\end{equation}
for every $\Phi \in \D(H_0)$.

By Lebesgue's Theorem, $(\tilde a_{1,j}^{\mu,m} + \tilde
a_{0,j}^{\mu,m} + \tilde b_j^{\mu,m})$ tends to zero when $m$
tends to zero. We then get, for every $\Phi \in \D(H_0)$,
\begin{equation}\label{eq:diff-zero}
 \lim_{m\mapsto 0} \|(H-H_m) \Phi\| = 0 .
\end{equation}
Thus the first statement of Theorem~\ref{thm3} is proved.

%\rightline{$\Box$}
\end{proof}
\subsection{Proof of (iii) of Theorem~\ref{thm3}}
The proof of \eqref{eq:48} will be a consequence of the following lemmas.

\begin{lem}\label{lem3.3}
For every $m \in (0,1)$, we have
\begin{equation}\nonumber
E_m :=\displaystyle\inf_{\scriptstyle \Phi \in \D(H_0)\atop
\scriptstyle \|\Phi\|=1} (H_m \Phi, \Phi) \leq 0
\end{equation}
and
\begin{equation}\label{eq2:lem3.3}
  \sup_{m\in(0,\,1)} |E_m| <\infty .
\end{equation}
\end{lem}

\begin{proof}
Since $\la a_\mu(k) \Omega_{ph},\Omega_{ph}\ra = \la\Omega_{ph},a_\mu^*(k)
\Omega_{ph}\ra = 0$, we have
 $$
  \la H_{I,m}^{(1)} \Omega_D \otimes \Omega_{ph},
  \Omega_D \otimes \Omega_{ph}\ra = 0 ,
 $$
and
 $$
  \la(\1 \otimes H_{ph})\, \Omega_D \otimes \Omega_{ph},
  \Omega_D \otimes\Omega_{ph}\ra = 0 .
 $$
Furthermore we also obviously have
 $$
  \la H_I^{(2)} \Omega_D,\, \Omega_D\ra = 0 .
 $$
Hence we get
\begin{eqnarray*}
\la H_m \Omega_D \otimes \Omega_{ph}, \Omega_D \otimes \Omega_{ph}\ra &=&
\la(d\Gamma(H_D) \otimes \1)\, \Omega_D \otimes \Omega_{ph}, \Omega_D \otimes
\Omega_{ph}\ra =0
\end{eqnarray*}
and
 $$
  E_m = \displaystyle \inf_{\scriptstyle
  \Phi \in \D(H_0)\atop \scriptstyle
  \|\Phi\|=1} \la H_m\Phi,\Phi\ra \leq
  \la H_m \Omega_D \otimes \Omega_{ph}, \Omega_D
  \otimes \Omega_{ph}\ra = 0
 $$
By Lemma~\ref{lem2.9}, which also holds when $v_i^\mu(k)$ is
replaced by $v_{i,m}^\mu(k)$, we get for some finite constant $C$
which does not depend on $m\in (0,1)$:
\begin{equation}\label{add61}
 \|H_{I,m}^{(1)} \Phi\| \leq C \left( \|H_0 + m_0 c^2)\Phi\| +
 \|\Phi\| \right)
\end{equation}
for every $\Phi\in\mathfrak{D}(H_0)$. From
Corollary~\ref{corr-quartic} we have, for some finite constant
$\tilde{C}$:
\begin{equation}\label{add62}
 \| (H_I^{(2)}\otimes \1) \Phi\| \leq \tilde{C} \left(
 \|(H_0 + m_0 c^2)\Phi\| + \| \Phi \|\right)
\end{equation}
for every $\Phi\in\mathfrak{D}(H_0)$. Thus \eqref{add61},
\eqref{add62} and the Kato-Rellich Theorem yield
\eqref{eq2:lem3.3}.
\end{proof}

In Section~\ref{S4.3} below, we will show that, for $|g_1| +
|g_2|$ sufficiently small, $H_m$ has a ground state $\Phi_m
(g_1,g_2)$, i.e. there exists, for every $m \in (0,1)$, a
normalized solution to $H_m \Phi_m(g_1,g_2) = E_m
\Phi_m(g_1,g_2)$.

We have
\begin{lem}\label{lem3.4}
There exist $\widetilde{g_0} >0$ and $\nu(\widetilde{g_0}) > 0$,
independent of $m\in (0,1)$, such that
 $$
  \|(H_0 + m_0 c^2) \Phi_m(g_1,g_2)\| \leq
  \nu(\widetilde{g_0}) ,
 $$
for every $m\in (0,1)$ and for every $(g_1,\,g_2)$ such that
$|g_1| + |g_2| \leq \widetilde{g_0}$.
\end{lem}

\begin{proof}
For simplicity, we will drop the dependence on $(g_1,g_2)$ in
$\Phi_m(g_1,g_2)$ by denoting $\Phi_m := \Phi_m(g_1,g_2)$. We have
\begin{equation}\label{eq:49}
\begin{split}
 H_0 \Phi_m = H_m \Phi_m - g_1\, H_{I,m}^{(1)} \Phi_m -g_2
 (H_I^{(2)}\otimes\1)\Phi_m \\
 = E_m \Phi_m - g_1\, H_{I,m}^{(1)}
 \Phi_m - g_2 (H_I^{(2)}\otimes\1)\Phi_m .
\end{split}
\end{equation}
>From \eqref{add61}, \eqref{add62} and \eqref{eq:49} we obtain
\begin{equation}\nonumber
 \|H_0 \Phi_m\| \leq |E_m| +\left(|g_1| C + |g_2| \tilde{C}\right)
 \left[\| (H_0 + m_0 c^2)
 \Phi_m\| + 1 \right] ,
\end{equation}
which implies
\begin{equation}\nonumber
 \| (H_0 + m_0 c^2) \Phi_m \|\leq \left(m_0 c^2 + |E_m| + |g_1|C +
 |g_2| \tilde{C}\right) \left( 1- |g_1|C - |g_2| \tilde{C}\right)^{-1}
\end{equation}
To conclude, we apply Lemma~\ref{lem3.3}, and we choose
$\widetilde{g_0} >0$ such that $|g_1| \, {C} + |g_2|\tilde{C} \leq
1/2$ for every $(g_1, g_2)$ such that $|g_1| + |g_2| \leq
\widetilde{g_0}$.
\end{proof}

\begin{lem}\label{lem3.5}
There exists $C>0$ such that
 $$
  \sum_{\mu =1,2} \|[1\otimes a_\mu(k), H_{I,m}^{(1)}]\,
  \Phi\|^2 \leq C \frac{G(k)^2}{E_0^2} \|( H_0 +m_0c^2)\Phi \|^2
 $$
for every $\Phi \in \D(H_0)$ and for a.e. $k \in \er^3$. Here
\begin{eqnarray*}
G(k)^2 &=&\!\!\! \sum_{\scriptstyle \mu =1,2} \,
\sum_{\gamma,\gamma',n,\ell}\!\! |G_{d,\gamma,\gamma',n,l}^\mu
(k)|^2 +4\sum_{\mu =1,2}\, \sum_{r=+,-} \sum_{\gamma,\gamma', n}
\int\! |G_{d,r,\gamma,\gamma',n}^\mu(p;k)|^2 \d  p\, \\
\noalign{\vskip 8pt} &+&4\sum_{\mu =1,2}\, \sum_{\gamma,\gamma'}
\int\int|G_{+,-,\gamma,\gamma'}(p,p';k)|^2\d  p\d  p'\,\\
&+&\sum_{\mu =1,2}\, \sum_{r=+,-}\sum_{\gamma,\gamma'}
\int\int|G_{r,r,\gamma,\gamma'}(p,p';k)|^2\d p\d  p'
\end{eqnarray*}
\end{lem}

\begin{proof}

By using the commutation relations, a simple computation shows that
\begin{eqnarray*}
&& [\1\otimes a_\mu(k),H_{I,m}^{(1)}] = \sum_{i=1}^6 v_{i,m}^\mu(k)
\end{eqnarray*}

The lemma now follows from Proposition~\ref{nprop3.4} and
Proposition~\ref{prop2.5}.
\end{proof}

\begin{lem}\label{lem3.6}
There exists $C > 0$, independent of $g_1$ and $g_2$ for $|g_1| +
|g_2| \leq \widetilde{g_0}$, such that
 $$
  (\1\otimes
  N_{ph} \Phi_m, \Phi_m) \leq C\ g_1^2\ \left( \int_{\er^3}
  \frac{G(k)^2}{\omega(k)^2}\, \d  ^3k \right),
 $$
for every $m \in (0,1)$. Here $$ N_{ph} := \sum_{\mu =1,2}
\int_{\er^3} \d ^3k\, a_\mu^*(k)\, a_\mu(k) $$ is the operator
number of photons.
\end{lem}

\begin{proof}

One easily checks that we have
 $$
  [\1\otimes a_\mu(k), d\Gamma(H_D) \otimes
  \1]\, \Psi = 0 ,
 $$
 $$
  [\1\otimes a_\mu(k), H_I^{(2)}\otimes\1]\Psi = 0 ,
 $$
and
 $$
  [\1\otimes a_\mu(k), \1 \otimes H_{ph}]\,
  \Psi = \omega(k)\,( \1 \otimes
  a_\mu(k))\, \Psi ,
 $$
for every $\Psi \in \D(H_0)$ and for a.e. $k \in \er^3$.

We then obtain :
 $$
  (H_m +\omega(k))\, \1\otimes a_\mu(k) - \1\otimes a_\mu(k)\,
   H_m = g_1[H_{I,m}^{(1)}, \1 \otimes a_\mu(k)] ,
 $$
for a.e. $k \in \er^3$.

Applying this equality to $\Phi_m$ and taking the scalar product with $\1
\otimes a_\mu(k)\, \Phi_m$, we get
\begin{equation}\label{eq:51}
\|\1\otimes a_\mu(k) \Phi_m\| \leq \displaystyle\frac{|g_1|}{\omega(k)}\
\|[\1\otimes a_\mu(k), H_{I,m}]\, \Phi_m\|
\end{equation}
for a.e. $k\in \er^3$.

Since
 $$
  \la\1\otimes N_{ph}\, \Phi_m,\Phi_m\ra = \sum_{\mu=1,2} \int \|\1\otimes
  a_\mu(k)\, \Phi_m\|^2 \d  ^3k
 $$
Lemma~\ref{lem3.6} follows from \eqref{eq:51}, Lemma~\ref{lem3.4} and
Lemma~\ref{lem3.5}.
\end{proof}

\begin{lem}\label{lem3.7}
Fix $\lambda$ in $(0, m_0 c_0^2)$. There exists $\delta_{g_1,g_2}(\lambda) > 0$
such that $\delta_{g_1,g_2}(\lambda)$ tends to zero when $|g_1|+|g_2|$ tends to
zero and
 $$
  \la P^\bot_D(\lambda) \otimes P_{\Omega_{ph}}\Phi_m, \Phi_m\ra \leq
  \delta_{g_1,g_2}(\lambda)
 $$
for every $m \in (0,1)$. Here $P^\bot_D(\lambda) =
\1-P_D(\lambda)=\1_{[\lambda,\ \infty)}(d\Gamma(H_D))$.
\end{lem}

\begin{proof}
Using $P_{\Omega_{ph}} H_{ph} = 0$, we get
\begin{eqnarray}
 \lefteqn{(P_D(\lambda)^\bot \otimes P_{\Omega_{ph}})\, (H_m - E_m)} & &
 \nonumber \\
 & = & P_D(\lambda)^\bot (d\Gamma(H_D)-E_m) \otimes P_{\Omega_{ph}} +
 g_1 (P_D(\lambda)^\bot \otimes P_{\Omega_{ph}})\, H_{I,m}^{(1)} \nonumber \\
 & & + g_2 \left( P_D(\lambda)^\perp\otimes P_{\Omega_ph}\right)
 H_I^{(2)}\otimes\1 . \label{eq:52}
\end{eqnarray}
>From \eqref{eq:52} we obtain
\begin{eqnarray}
 0 &=& (P_D^\bot (\lambda) \otimes P_{\Omega_{ph}})\, (H_m-E_m)\, \Phi_m
 \label{eq:53} \\ \noalign{\vskip 8pt}
 &=& P_D^\bot (\lambda)\, (d\Gamma(H_D)-E_m) \otimes
 P_{\Omega_{ph}}\Phi_m +g_1\ P_D^\bot (\lambda) \otimes
 P_{\Omega_{ph}}H_{I,m}^{(1)}\Phi_m \nonumber\\
 & & + g_2 \left( P_D(\lambda)^\perp\otimes P_{\Omega_{ph}}\right)
 H_I^{(2)}\otimes\1 .\nonumber
\end{eqnarray}
Since
 $$
  d\Gamma(H_D) P_D (\lambda)^\bot \geq \lambda\ P_D^\bot
  (\lambda) ,
 $$
by \eqref{eq:53} and Lemma~\ref{lem3.3}, we obtain
\begin{equation}\nonumber
\begin{split}
  {\la P_D^\bot(\lambda)\otimes
  P_{\Omega_{ph}} \Phi_m, \Phi_m\ra }\leq  &  -\lambda ^{-1}
  \bigg[g_1\Big\langle(P_D^\bot(\lambda)\otimes
  P_{\Omega_{ph}}) H_{I,m}^{(1)} \Phi_m, \Phi_m\Big\rangle \\
  & + g_2\Big\langle(P_D^\bot(\lambda)\otimes
  P_{\Omega_{ph}}) (H_{I}^{(2)}\otimes\1) \Phi_m, \Phi_m\Big\rangle\bigg] ,
\end{split}
\end{equation}
for every $m\in (0,1)$.

Lemma~\ref{lem3.7} then follows by remarking that
Corollary~\ref{corr2.10} also holds for $H_{I,m}^{(1)}$ and by
using Lemma~\ref{corr-quartic} and Lemma~\ref{lem3.4}.
\end{proof}

We are now able to prove the inequality \eqref{eq:48}, i.e., point (iii) of
Theorem~\ref{thm3}.

\begin{proof}
The equation \eqref{eq:48} is equivalent to
 $$
  \la(P_D^\bot(\lambda) \otimes
  P_{\Omega_{ph}} + \1 \otimes P^\bot_{\Omega_{ph}})\, \Phi_m,\Phi_m\ra \leq
  \delta_{g_1,g_2},
 $$
Since $P_{\Omega_{ph}}^\bot\leq N_{ph},$ it suffices to show that
 $$
  \la(P_D^\bot(\lambda) \otimes P_{\Omega_{ph}} + \1\otimes N_{ph})\Phi_m,
  \Phi_m\ra\leq \delta_{g_1,g_2},
 $$
which follows from Lemma~\ref{lem3.6} and Lemma~\ref{lem3.7}.

>From the fact that $\delta_{g_1, g_2}$ tends to zero when $|g_1| +
|g_2|$ tends to zero, we deduce the existence of
$\widetilde{g_{00}}$ such that $\delta_{g_1, g_2} <1$ for $|g_1| +
|g_2| \leq \widetilde{g_{00}}$.
\end{proof}

\subsection{Proof of (ii) of Theorem~\ref{thm3}:
Existence of a ground state for $H_m$\label{S4.3}}

>From now on we mimic the proof in \cite{ref2} and \cite{ref4} (See
also \cite{H1}, \cite{H2}, \cite{ref15} and \cite{ref19}) and we use
the same notations as in \cite{ref3}.

For $m > 0$, we set
 $$
  {\mathcal A}_\ell= \{ k \in \er^3\ ; \ \omega(k) \geq
  m\}, \quad {\mathcal A}_s= \er^3\setminus  {\mathcal A}_\ell ,
 $$
and
 $$
  h_{s/\ell} =
  L^2({\mathcal A}_{s/\ell}) .
 $$
Let ${\mathfrak F}(h_{\ell/s})$ be the bosonic Fock spaces associated with
$h_{\ell/s}$, with associated vacua $\Omega_{\ell / s}$.

Using the following lemma (See \cite{ref4} and \cite{DG}),

\begin{lem}\label{lem3.8}
Let $h_i$, $i = 1,2$ be two Hilbert spaces. There exists a unitary operator
from the photonic Fock space over $h_1
\oplus h_2$ to
 $$
  {\mathfrak F}_{ph}(h_1) \otimes {\mathfrak F}_{ph}(h_2).
 $$
\end{lem}
\vspace{0.5cm}\noindent We may identify $H_{ph}$ (resp.
$g_1H_I^{(1)}$) with
 $$
  H_{ph,\ell} \otimes \1_s + \1_\ell \otimes H_{ph,s}, \ \
  (\mbox{resp.}\ g_1 H_{I,m}^{(1)}
  \otimes \1_s).
 $$
Here
 $$
  H_{ph,\ell/s} = \sum_{\mu =1,2} \int_{{\mathcal A}_{\ell/s}} \d  ^3k\,
  \omega(k)\, a_\mu^*(k)\, a_\mu(k).
 $$
We set
 $$
  H_{m,\ell} =d\Gamma( H_D)
  \otimes \1_\ell + \1_D \otimes H_{ph,\ell} + g_1\, H_{I,m}^{(1)}
  + g_2 H_I^{(2)}\otimes\1_\ell .
 $$
In this representation, $H_m$ appears as
\begin{equation}\label{eq:54}
H_m = H_{m,\ell} \otimes \1_s + \1_D \otimes \1_\ell \otimes H_{ph,s} .
\end{equation}
We then have
\begin{lem}\label{lem3.9}
$H_{m,\ell}$ has a ground state $\Phi_{m,\ell}$ if and only if $H_m$ has a
ground state $\Phi_m = \Phi_{m,\ell} \otimes \Omega_s$.
\end{lem}

\begin{proof}

In view of \eqref{eq:54}, one has
\begin{equation}\label{eq:55}
H_m(\Phi_{m,\ell} \otimes \Omega_s) = H_{m,\ell}\, \Phi_{m,\ell} \otimes
\Omega_s
\end{equation}
Furthermore, it results from \eqref{eq:54} that
\begin{eqnarray*}
\sigma(H_m) &=& \overline{\sigma(H_{m,\ell}) + \sigma(H_{ph,s}})\ =\  [\inf
\sigma(H_{m,\ell}), + \infty)
\end{eqnarray*}
which, together with \eqref{eq:55}, gives the lemma.
\end{proof}

By Lemma~\ref{lem3.9}, Theorem~\ref{thm3} then follows from

\begin{theo}\label{thm4}
There exists $\widetilde{g_{000}}$ such that, for every $(g_1,
g_2)$ such that $|g_1| + |g_2| \leq \widetilde{g_{000}}$ and $m
\in (0,1)$, $H_{m,\ell}$ has a ground state $\Phi_{m,\ell}$ at
$E_m = \inf \sigma(H_m) = \inf \sigma (H_{m,\ell})$.
\end{theo}

\begin{proof}
Let $\varepsilon > 0$ be a parameter. We decompose $\er^3$ into a
disjoint union of cubes of side lenght $\varepsilon$, $\er^3 =
\bigcup_{n\in (\varepsilon \zed^3)}\, n+ Q_\varepsilon$, where
$Q_\varepsilon = [-\frac{\varepsilon}{2},
\frac{\varepsilon}{2})^3$. As in \cite{ref3} and \cite{ref19}, for
$F \in L_{loc}^1(\er^3)$, we define its $\varepsilon$-average by
 $$
  \langle F \rangle_\varepsilon (k) =
  \varepsilon^{-3} \int_{n(k) +
  Q_\varepsilon}
  F(k')\, \d  ^3k'  ,
 $$
where $n(k)\in(\varepsilon\zed)^3$ is such that $k-n(k) \in
Q_\varepsilon$.

More generally, when $G \in L^2(\Sigma\times \er_k^3)$, we define its
$\varepsilon$-average with respect to $k$ by
 $$
  \langle G \rangle_\varepsilon
  (k) := G_\varepsilon(\cdot,k) =
  \varepsilon^{-3} \int_{n(k) +
  Q_\varepsilon}
  G(\cdot,k')\, \d  ^3k'  .
 $$
Let $H_{I,m}^\varepsilon$ be the operator obtained from
$H_{I,m}^{(1)}$ by substituting $\langle
v_{i,m}^\mu\rangle_\varepsilon (k)$ for $v_{i,m}^\mu(k)$.

Similarly, we define $H_{ph,\ell/s}^{\varepsilon}$ by substituting
$\langle \omega\rangle_\epsilon (k)$ for $\omega(k)$ in
$H_{ph,\ell/s}$.

A simple calculation shows that
 $$
  |\omega(k) - \langle \omega
  \rangle_\varepsilon(k)| \leq \tilde C \, \varepsilon\, \omega(k) ,
 $$
It implies
\begin{equation}\label{eq:starr}
  \pm (H_{ph,\ell} - H_{ph,\ell}^{\varepsilon}) \leq \tilde
  C\, \varepsilon\, H_{ph,\ell}.
\end{equation}
Set $H_{m,\ell}^\varepsilon = d\Gamma(H_D)\otimes\1_\ell + \1_D \otimes
H_{ph,\ell}^\varepsilon + g_1 H_{I,m}^\varepsilon + g_2
H_I^{(2)}\otimes\1_l$. Combining \eqref{eq:starr} with
Corollary~\ref{corr2.10} and using the fact that
 $$
  H_{m,\ell}^\varepsilon - H_{m,\ell} = \1_D\otimes H_{ph,\ell}^{\varepsilon} -
  \1_D\otimes H_{ph,\ell} + g_1(H_{I,m}^\varepsilon - H_{I,m}^{(1)}),
 $$
we obtain
\begin{equation}\label{eq:56}
 \|(H_{m,\ell}^\varepsilon - H_{m,\ell})\, (H_{m,\ell} - E_m + 1)^{-1}\| \leq
 {\rm const.}  \left(\varepsilon + \sum_{\beta=0,1}\, \sum_{j=1}^6\,
 (a_{\beta,j}^{\mu,m,\varepsilon} +
 b_j^{\mu,m,\varepsilon})\right),
\end{equation}
where $a_{\beta,j}^{\mu,m,\varepsilon}$ (resp. $b_j^{\mu,m,\varepsilon})$
denotes the expression \eqref{eq:41} (resp. \eqref{eq:42}) that we obtain when
we substitute $v_{j,m}^\mu(k) - \langle v_{j,m}^\mu\rangle_\v (k)$ for
$v_j^\mu(k)$.

Now, the right hand side of \eqref{eq:56} converges to zero as
$\varepsilon \mapsto 0$.

On the other hand, by mimicking \cite{ref3} and \cite{ref4} one
shows that there exists $\widetilde{g_{000}}$ such that the finite
volume approximation $H_{m,\ell}^\varepsilon$ has discrete
spectrum in $(-\infty, E_m + m_2)$, for $m_2 < m$ and for $(g_1,
g_2)$ such that $|g_1| + |g_2| \leq \widetilde{g_{000}}$.

Theorem~\ref{thm4} then follows from \eqref{eq:56} and from the
following lemma (see \cite{HKS}).
\end{proof}

\begin{lem}\label{lem3.11}
\quad Let $(T_n)_{n\geq 1}$ and $T$ be bounded below self-adjoint
operators on a Hilbert space. Suppose that
\begin{itemize}
\item[i)] $T_n \rightarrow T$ as $n \mapsto + \infty$ in the norm
resolvent sense.
\item[ii)] $T_n$ has purely discrete spectrum in $[\inf \sigma(T_n),
\inf \sigma(T_n) + C)$, where $C$ is a constant independent of $n$.
\end{itemize}
Then $T$ has purely discrete spectrum in $[\inf \sigma(T), \inf
\sigma(T) + C)$.
\end{lem}

\noindent {\it Proofs of Theorem~\ref{thm2} and Theorem~\ref{thm3}.}
Choosing $g_0 = \min \{\widetilde{g_{0}}, \widetilde{g_{00}},
\widetilde{g_{000}}\}$, Theorem~\ref{thm3} follows from
\eqref{eq:diff-zero}, Lemma~\ref{lem3.7} and Theorem~\ref{thm4}. The
proof of Theorem~\ref{thm2} is easily deduced from Theorem~\ref{thm3}
as it is explained at the beginning of the Section~\ref{section4}.
%%%%%%%%%%%%%%%%%%%%%%%%%%%%%%%%%%%%%%%%%%%%%%%%%%%%%%%%
%%%%%%%%%%%%%%%%%%%%  APPENDIX  %%%%%%%%%%%%%%%%%%%%%%%%
%%%%%%%%%%%%%%%%%%%%%%%%%%%%%%%%%%%%%%%%%%%%%%%%%%%%%%%%
\section{Appendix}
\subsection{Proof of \eqref{eq:45} and \eqref{eq:46} }

We keep the same notations as in Section~\ref{S3}.

\bigskip
\begin{lem}\label{lema1}
Under the assumptions of Theorem~\ref{thm1}, we have
\begin{equation}\label{eq:a1}
  \| \int \d  ^3k\, v_i^\mu(k)^*
  \otimes a_\mu(k)\, \Psi\| \leq b_i^\mu \|(N_D+1)^{1/2} \otimes H_{ph}^{1/2}
  \Psi\| ,
\end{equation}
and
\begin{eqnarray}\label{eq:a2}
 \| \int \d  ^3k\, v_i^\mu(k) \otimes a_\mu^*(k)\, \Psi\|^2
 & \leq & (a_{1,i}^\mu)^2 \|(N_D+1)^{1/2} \otimes H_{ph}^{1/2} \Psi\|^2
 \nonumber \\
 & & + (a_{0,i}^\mu)^2 \left[\varepsilon \|(N_D+1) \otimes \1\, \Psi\|^2 +
 \frac{1}{4\varepsilon} \|\Psi\|^2 \right],
\end{eqnarray}
for every $\Psi\in \D(H_0)$ and every $\varepsilon > 0$.
\end{lem}
\bigskip

\begin{proof}

We only give the proof of \eqref{eq:a2} for $i=5$. The case of $i=6$ is quite
similar.

The other cases, i.e. $i\not= 5,6$ are simpler since $v_i^\mu(k)$ is bounded on
${\mathfrak F}_D$ for $i\not= 5,6$.

In the following we omit the indexes $\mu$ and 5.

Set
 $$
  a(v)  = \int \d  ^3k\, v^*(k) \otimes a(k)\\
 $$
\begin{equation}\nonumber
a(v)^* = \int \d  ^3k\, v(k) \otimes a^*(k)
\end{equation}
Let $u \in \D(H_0)$. For $k \in \er^3$, set
 $$
  \Phi(k) =
  \omega(k)^{1/2}((N_D+1)^{1/2} \otimes a(k))\, u.
 $$
One has
\begin{eqnarray}
 \int \|\Phi(k)\|^2 \d  ^3k &=& \int \omega(k)\,\left\la
 (N_D+1)^{1/2}\otimes a(k) u,
 (N_D+1)^{1/2} \otimes a(k)u\right\ra \d  ^3k\nonumber\\
 \noalign{\vskip 8pt} &=& \left\la
 (N_D+1)^{1/2} \otimes \int \omega(k) a^*(k) a(k)
 \d  ^3k u, (N_D+1)^{1/2} \otimes \1\ u  \right\ra\nonumber\\
 \noalign{\vskip 8pt} &=& \left\la((N_D+1)^{1/2} \otimes
 H_{ph})\, u, ((N_D+1)^{1/2} \otimes \1)\, u\right\ra\nonumber\\
 \noalign{\vskip 8pt} &=&
 \|((N_D+1)^{1/2} \otimes H_{ph}^{1/2})u\|^2 \label{eq:a4}
\end{eqnarray}
Using \eqref{eq:a4}, we get
\begin{eqnarray*}
 \|a(v) u\|^2
 &\!\!\!\!\! = &\!\!\!\!\! \int \left\la v^*(k) \otimes a(k) u,
 v^*(k') \otimes a(k') u\right\ra \d  ^3k\,
 \d  ^3k'\\
 &\!\!\!\!\! = &\!\!\!\!\! \int \omega(k)^{-\frac{1}{2}}
 \omega(k')^{-\frac{1}{2}}\\
 &\!\!\!\!\! &\!\!\!\!\! \times
 \Big\la v^*(k)((N_D+1)^{-\frac12}\!\otimes\! \1) \Phi(k),v^*(k')
 ((N_D+1)^{-\frac{1}{2}}\!\otimes\! \1) \Phi(k')\Big\ra \d  ^3k \, \d  ^3k' \\
 &\!\!\!\!\! \leq &\!\!\!\!\! \left[\int
 \omega(k)^{-\frac{1}{2}} \|v^*(k)
 (N_D+1)^{-\frac{1}{2}}\|_{{\mathfrak F}_D} \|\Phi(k)\|\, \d  ^3 k\right]^2.
\end{eqnarray*}
Cauchy-Schwarz inequality now implies that $$ \|a(v) u\|^2 \leq
\left( \int \omega(k)^{-1} \|v^*(k)\, (N_D+1)^{-1/2}\|_{{\mathfrak
F}_D}^2 \d  ^3k\right)\, \left( \int \|\Phi(k)\|^2 \d  ^3
k\right), $$ which, together with \eqref{eq:a4} gives
\eqref{eq:a1}.

Let us now prove \eqref{eq:a2}. We have
 $$
  \|a^*(v) u\|^2 = \int \left\la v(k) \otimes
  a^*(k) u, v(k') \otimes a^*(k')u\right\ra\d  ^3k\, \d  ^3k' .
 $$
Using the commutation relations for $a$ and $a^*$ we obtain
\begin{eqnarray}
\|a^*(v) u\|^2 & = & \int\left\la v(k) \otimes a^*(k) a(k') u, v(k') \otimes
\1\, u\right\ra  \d  ^3k\, \d  ^3k'\nonumber \\ & & + \int \|(v(k) \otimes
\1)\, u\|^2 \d ^3k \label{eq:a5}
\end{eqnarray}
The first term in the right hand side of \eqref{eq:a5} can be written in the
following way
\begin{eqnarray*}
 \lefteqn{\int \left\la
 v(k) \otimes a(k') u, v(k') \otimes a(k) u\right\ra\d  ^3k\,
 \d ^3k'} & &
 \\ & = & \int \omega(k)^{-\frac12} \omega(k')^{-\frac12}\\ & &
 \times \Big\la v(k)\, \big((N_D+1)^{-\frac12} \otimes \1\big) \Phi(k'),
 v(k')\big((N_D+1)^{-\frac12} \otimes \1\big) \Phi(k)\Big\ra \d  ^3k\, \d  ^3k'
\end{eqnarray*}
Hence
\begin{eqnarray*}
 \lefteqn{ \int\left\la
 v(k) \otimes a^*(k) a(k') u, v(k') \otimes \1\, u\right\ra \d
 ^3k\, \d ^3k'} & &\\ & \leq & \left[ \int \omega(k)^{-1/2} \|v(k)(N_D+1)
 ^{-1/2}\|_{{\mathfrak F}_D} \|\Phi(k)\| \d  ^3k\right]^2\\ \noalign{\vskip 8pt}
 & \leq & \left( \int \omega(k)^{-1} \|v(k) (N_D+1)^{-1/2}\|_{{\mathfrak F}_D}^2
 \d  ^3k\right)\, \left( \int \|\Phi(k)\|^2 \d  ^3k \right)
\end{eqnarray*}
which, together with \eqref{eq:a4}, yields
\begin{eqnarray}
 \lefteqn{ \int\left\la
 v(k) \otimes a^*(k) a(k') u, v(k') \otimes \1\, u\right\ra \d  ^3k\,
 \d  ^3k'} & & \nonumber \\
 & \leq & \left( \int \omega(k)^{-1/2} \|v(k)
 (N_D+1)^{-1/2}\|_{{\mathfrak F}_D}^2 \d  ^3k\right]^2 \, \left\|\left((N_D+1)^{1/2} \otimes
 H_{ph}^{1/2}\right) u\right\|^2\nonumber \\
 & = & \left( \int \omega(k)^{-1} \|
 (N_D+1)^{-1/2} v(k)^*\|_{{\mathfrak F}_D}^2 \d  ^3k\right)\, \left\|\left((N_D+1)^{1/2}
 \otimes H_{ph}^{1/2}\right) u\right\|^2 .\label{eq:a6}
\end{eqnarray}
For the second term in the right hand side of \eqref{eq:a5}, we write
\begin{eqnarray*}
 \lefteqn{ \int \|(v(k) \otimes \1)\, u\|^2 \d  ^3k } & & \\
 & \leq & \left( \int \|v(k) (N_D+1)^{-\frac12}\|_{{\mathcal
 F}_D}^2 \d  ^3k \right) \| ((N_D+1)^{\frac12}\otimes \1)u\|^2 \\
 &=& \left( \int \|v(k) (N_D+1)^{-\frac12}\|_{{\mathfrak F}_D}^2 \d  ^3k \right)
 \left((N_D+1) u,\, u\right) .
\end{eqnarray*}
Using for all $\varepsilon>0$, $a b \leq {\varepsilon}\, a^2 +
\frac{1}{4\varepsilon}\, b^2$, we get

\begin{eqnarray}
\lefteqn{\int \|(v(k) \otimes \1)\, u\|^2 \d  ^3k } & & \nonumber \\ & \leq &
\left( \int \|v(k) (N_D+1)^{-1/2}\|^2 \d  ^3k\right)\, \left[ \varepsilon
\|(N_D+1) u\|^2 + \frac{1}{4\varepsilon}\, \|u\|^2 \right]\label{eq:a7}
\end{eqnarray}
Finally, \eqref{eq:a5}, \eqref{eq:a6} and \eqref{eq:a7} give \eqref{eq:a2}.
\end{proof}

\newpage

\bibliographystyle{amsalpha}

\end{document}